\documentclass[fleqn,usenatbib]{mnras}

\usepackage{newtxtext,newtxmath}

\usepackage[T1]{fontenc}

\DeclareRobustCommand{\VAN}[3]{#2}
\let\VANthebibliography\thebibliography
\def\thebibliography{\DeclareRobustCommand{\VAN}[3]{##3}\VANthebibliography}

\usepackage{graphicx}	
\usepackage{amsmath}
\usepackage{float}

\usepackage[todonotes={textsize=footnotesize}]{changes} 
\definechangesauthor[name=Jorge,color=red]{JZ}

\usepackage{geometry}
\title{Concentration of water in the lower Mars' atmosphere across solar cycle}

\author[Molina Córdoba et al.]{
Johan  Nicolás  Molina Córdoba$^{1}$\thanks{E-mail: jomolinac@unal.edu.co}, Santiago Vargas Dom\'inguez $^{1}$\thanks{E-mail: svargasd@unal.edu.co}, Jorge I. Zuluaga$^{2}$\thanks{E-mail: jorge.zuluaga@udea.edu.co}, 
\\
$^{1}$Universidad Nacional de Colombia, Observatorio Astronómico Nacional, Carrera 45 No. 26-85. Ed. 413 Bogotá, Colombia\\
$^{2}$SEAP/FACom, Instituto de F\'{\i}sica - FCEN, Universidad de Antioquia, Calle 70 No. 52-21, Medell\'in, Colombia\\
}

\date{Accepted XXX. Received YYY; in original form ZZZ}

\pubyear{2023}

\begin{document}
\label{firstpage}
\pagerange{\pageref{firstpage}--\pageref{lastpage}}
\maketitle

\begin{abstract}\\

Mars' thin, CO$_2$-rich atmosphere poses a unique puzzle involving composition, climate history, and habitability. This work explores the intrincate relationship between Mars' atmospheric variations and dynamic solar activity patterns. We focus on periodic oscillations in H$_2$O vapor and the Pectinton solar flux index in the $\lambda$ = 10.7 cm radio band, around the characteristic 11-year solar cycle. Periodic Mars activity was studied using data from Mars Express' SPICAM instrument spanning 2004-2018. The Lomb-Scargle Periodogram method was applied to analyze the power spectra of both signals around this period, calibrated using peaks associated with the seasonal Martian cycle. This method was validated by analyzing power spectra of chemical species abundances in Earth's atmosphere, obtained from the NRLMSISE 00 empirical model provided by the National Oceanic and Atmospheric Administration (NOAA). Model executions reproduced chemical abundance data for various atmospheric species (N$_2$, O$_2$, N, H$_2$, Ar, and He) at two reference heights (upper mesosphere and low ionosphere) over a 1961-2021 time span. Results suggest a connection between variability in H$_2$O vapor concentration in Mars' atmosphere and fluctuations in the Pectinton solar flux index. We propose the Lomb-Scargle Periodogram method as a heuristic for studying oscillatory activity in planetary atmospheres with non-uniformly sampled data. While our results provide valuable insights, further analysis, cross-referencing with data from different orbiters, is required to deepen our understanding of these findings in the fields of planetary climatology and atmospheric physics.

\end{abstract}




\section{Introduction}

\subsection{Characterization of the Martian Atmosphere}

Within the research realm of planetary atmospheres, the use of spacecraft orbiting different planets in our solar system (including Earth) since the 1960s has been essential to identifying atmospheric changes related to various phenomena. These phenomena include seasonal cycles (see eg. \citealt{Mars_seasonal_2002, jakosky1992seasonal_H2O, james1992seasonal}), magnetic activity and ionization levels in the ionosphere \citep{Tian_2022_magnetic_Mars}, the latter produced by incident ultraviolet radiation from the Sun.

The case of Mars is quite extensive, as it is the most explored planet within our planetary neighborhood after Earth. Currently, there are over 15 orbiter missions that have successfully reached the Martian atmosphere, not counting the landers that have touched down on its surface. Among these Mars orbiters, significant mention should be given in historical order to: Mariner, Mars Climate Orbiter (MCO), Mars Express, Mars Reconnaissance Orbiter (MRO), and Mars Atmosphere and Volatile Evolution (MAVEN). Each of these missions focuses on specific aspects of Mars' behavior. For instance, probes like MRO and MCO study wind speeds at different altitudes above the Martian surface, sense pressures and temperatures at various points in the atmosphere, as well as the atmospheric density at different heights. They also make estimates regarding dust abundance and ice distribution. Others, such as MAVEN and Mars Express, conduct surveys on the distribution of various atmospheric gases—H$_2$O, CO$_2$, Ar, He, O, O$_2$, H, N$_2$. 

MAVEN leads the current field of \textit{in situ} research on the Martian atmosphere and is designed especially to study the evolution and dynamics of the Martian atmosphere \citep{MAVEN_jakosky2015mars}. MAVEN works with data from the upper atmosphere, observing the loss of atmospheric gases due to photoionization processes and Jeans escape mechanisms. It examines the impact of Solar Energetic Particles (SEPs) on ion exchange processes at different altitudes in the upper Martian atmosphere, as well as their escape \citep{jakosky2018loss_Jeans}.

Within current models for studying and characterizing the atmosphere of Mars, it has been observed that similar to Earth's atmosphere, the upper layer of the Martian atmosphere is primarily ionized due to ultraviolet radiation from the Sun (see eg. \citealt{Tian_2022_magnetic_Mars} and reference there in). The Mars' ionosphere encompasses roughly 120 km from the outer vertical column of the Martian atmosphere; there, high ion densities are found \citep{Haider_Mars_ionosf_2011}, generated by the constant exposure to high-energy radiation emitted from the solar surface. 


On the other hand, atmosphere has a well-mixed region called the homosphere, a layer extending from the surface up to the homopause at around 120 km altitude.
Another region of interest is the thermosphere, where gases diffusively separate into individual chemical species following their scale heights. On Mars, this has an average value of around 11 km. On Earth, the thermosphere extends from 80 to 500 km in altitude, whereas on Mars, it spans from 100 to 230 km in altitude. In the Martian thermosphere, temperature increases with altitude and varies seasonally based on the planet's position in its orbit around the Sun. During the day, temperatures in the upper layer of the thermosphere can vary from 175 to 240 K, depending on Mars' orbital position at its aphelion and perihelion, respectively \citep{jain2023_mars_temperature}. The thermosphere merges with the exosphere, where lighter gases can become energized to attain escape velocities. Typically, this loss process begins around 220 km altitude, known as the exobase, where scale heights and mean free paths are comparable \citep{lee2015hot_scape_mars}. The dynamics of this region are driven by energy fluxes, planetary rotational momentum, and tidal waves propagating through the lower atmosphere. Effects from extreme ultraviolet (EUV) solar radiation introduce energetic particles in the lower atmosphere through flares and variable solar wind plasma flow \citep{MarsAtmosphereLayer}.

On the other hand, the study of Mars has been sequentially stimulated by the progressive investigation of solar activity and dynamics, and its influence on atmospheric changes in certain planets. For instance, studies like \cite{nagaraja2021Mars_solar} and \cite{GW_planetary} examine significant changes mapped in the Earth's and Martian atmospheres during particular moments of the known solar activity cycle. In \cite{nagaraja2021Mars_solar}, the authors analyze variations in atmospheric density based on solar wind flux for the first two weeks of June 2018 \footnote{There are indications, \cite{venkateswara2020_2018_event}, that during the first week of this month, there was an intense solar wind, evident in atmospheric variations on Mars.}, chosen according to the characteristic solar wind of that period. The results of this study \cite{nagaraja2021Mars_solar} demonstrate that the energetic plasma from the solar wind affected the concentrations of CO$_2$, O, Ar, and He, but not the concentrations of N$_2$ and CO, whose density values remained nearly constant. 

In the Martian atmosphere's Thermosphere/Exosphere regions, the densities of CO$_2$ and Ar exhibited decreasing trends simultaneously with decreases in the kinetic energy and plasma density of the solar wind. The opposite occurred for O and He. The authors indicate interpreting these novel findings by considering the role of a large number of dissociation and ionization reactions initiated by electron impact from the solar wind plasma.

During a solar cycle, ultraviolet (UV) and extreme ultraviolet (EUV) radiation impact planetary thermospheres through processes of photon absorption and atmospheric ionization. These effects bring about changes in the density and temperature of the thermosphere, notably. In \cite{GW_planetary}, the analysis is conducted for Solar Cycle 24, which has been characterized by an extended period of low solar activity with an unprecedented solar minimum \citep{danilov2020_24_solar_min}. From the perspective of atmospheric coupling, it is valuable to study the solar atmosphere during a solar minimum to identify changes in the propagation of effects from the lower atmospheres, caused, for example, by wind flows in the lower parts of the atmosphere. In this context, the authors conduct an atmospheric study of Mars in such a way that solar wind has the least impact on Martian atmospheric dynamics, achieving a parameterized state of the Martian atmosphere free from solar variability effects. Changes in the propagation of effects from the lower atmosphere are characterized as atmospheric gravity waves (GW), a concept supported by a scientific tradition within planetary sciences \citep{Zurek_WG_thermosphere_MAVEN, Thiemann_2018_ionosphere_solarFlare, medvedev2019gravity_MCG, astafyeva2019ionosphericWG}.

\subsection{Water in Martian Atmosphere}

When solar radiation penetrates the Martian atmosphere, it undergoes a series of interactions with the molecules present, giving rise to different physical processes that can affect how the radiation scatters and is absorbed. One of the key components interacting with solar radiation in the Martian atmosphere is carbon dioxide (CO$_2$). Since Mars' atmosphere is dominated by 95\% CO$_2$, this gas is primarily responsible for the radiative transfer mechanisms on the planet \citep{franz2017initial_CO2_abund_Mars}. Carbon dioxide is a greenhouse gas, meaning it has the ability to absorb solar radiation and trap heat within the atmosphere. This is crucial for understanding the climate on Mars and how it has changed over time. The energy-absorbing property of CO$_2$ explains, for instance, why, even in an overall cold atmosphere (average temperature of the Martian atmosphere), there are regions on Mars with varying degrees of water vapor abundance.

Despite the Martian atmosphere being extremely thin and dry, there are still small amounts of water vapor (H$_2$O) present (0.01\% of the atmosphere). While CO$_2$ is commonly attributed as the main contributor, it's important to note that the low pressures in Mars' atmosphere (ranging from 0.0007 to 0.0009 atmospheres) cause vaporization temperatures on Mars to fall between 0 and 5$^{\circ}$C \citep{jakosky1992seasonal_H2O}, in addition to driving a hydrological cycle characterized by exchange only between the solid and gaseous phases.

The initial findings concerning the vertical distribution of water vapor in 22 orbits of the Mars Express mission were published by \cite{fedorova2009solaroccult}, along with altitude profiles of CO$_2$ distribution and some aerosols. This work inferred the vertical distribution of aerosols and H$_2$O density in a cloud layer in the Martian atmosphere from optically traceable properties in spectra acquired through solar occultations \citep[see][]{maltagliati2013occultations}. A significant discrepancy was found in this study between the water abundance above 30 km compared to estimates from the Mars Climate Database. The latter is based on classical atmospheric models of global circulation. On the other hand, water vapor has the capacity to absorb radiation in specific bands of the electromagnetic spectrum, which is important for studying its presence and distribution in the Martian atmosphere. It undergoes three types of energy transitions (rotational, electronic, and vibrational) that give rise to electromagnetic radiation absorption processes \citep{2009iupac_H2O_transition}. Vibrational transitions, for which electromagnetic energy absorption occurs in the mid and near-infrared region, particularly the $\mu$ band at 6 $\mu$m and the X band at 2.9 $\mu$m, are the specific bands studied to identify water vapor abundances in the Martian atmosphere. These transitions represent a fraction of the lines derived from synthetic spectra in the HITRAN database \citep{rothman2021history_HITRAN}, of which the line of special interest for Martian atmosphere study is the $\Psi$ line at 1.38 $\mu$m\footnote{This is the assigned code for the line of interest in Atmospheric Radiative Transfer Codes: a database of atmospheric models.}. These transitions occur at ambient temperature due to the internal energy of the covalent bonds within the water molecule, and they are the result of the continuous redistribution of charges within the molecule \citep{polyansky2003_vibrational_H2O}.

In general, the HITRAN spectroscopic database lists more than 37 thousand spectral lines for gaseous H$_2$O, spanning from UV to radio waves \citep{rothman2021history_HITRAN}. This database, based on radiative transfer models, is a fundamental resource for generating synthetic spectra for Mars, which are then compared with observations carried out by the IR channel of the SPICAM (Spectroscopy for the Investigation of the Characteristics of the Atmosphere of Mars) instrument on Mars Express. This comparison aims to assess the level of uncertainty in detections made by this and other orbiters.\\
As previously mentioned, H$_2$O vapor can absorb and re-emit infrared radiation, thus trapping heat in the atmosphere and warming the planet's surface. However, the amount of water vapor in the Martian atmosphere is highly variable both spatially and temporally, closely tied to the planet's water cycle and its interaction with the surface and subsurface \citep{davies1981mars_H2O_cycle}, Mars' orbital period \citep{jakosky1992seasonal_H2O}, and presumably the solar activity cycle.

Finally, in addition to atmospheric gases, the presence of dust in the Martian atmosphere also plays a significant role in radiation transport mechanisms. Dust in the atmosphere can absorb and scatter radiation, which can impact the amount of radiation reaching the planet's surface. Dust can also influence temperature and the amount of sunlight penetrating the atmosphere, subsequently affecting Mars' climate \citep{forget1998improved_dust_Mars}.

\subsection{Space Weather: The Earth-Sun case}

Previously, studies were referenced concerning the atmospheric behavior of Mars during specific phases of the solar cycle, both at its maximum in the work of \cite{nagaraja2021Mars_solar}, accounting for changes in the concentration of various chemical species, and at its minimum in \cite{GW_planetary}, addressing the internal behavior of Mars' atmosphere through GW. In this section, emphasis is placed on the context of variability in atmospheric observables over complete periods of solar activity spanning time ranges exceeding 11 years (solar activity period).

The work we conducted is strongly influenced by the research \cite{Terminator_Corr_Earth}, where the correlation between a characteristic solar event known as the terminator (the end of Hale magnetic cycles) and major oscillations in Earth's oceanic indices determining transient climatic effects in the tropical Pacific, such as El Niño and La Niña phenomena, is studied with high statistical significance. In \cite{Terminator_Corr_Earth}, data from the last 60 years of solar photosphere observations are recorded to establish a robust link between Earth's troposphere and the end of the solar cycle known as the terminator\footnote{A terminator is the event that marks the transition from one sunspot cycle to the next. It is an abrupt event occurring at the solar equator resulting from the annihilation/cancellation of opposite-polarity magnetic activity bands in the heart of the 22-year solar cycle; in other words, the old-cycle flux nulls out on the disk. \cite{mcintosh2021deciphering_hale}}. Radiative proxies of solar activity are also observed at other wavelengths, such as X-rays from 1 to 8 \AA, measured by the GOES family of spacecraft (e), and the indicator of solar chromospheric variability measured through the ultraviolet emission of individually ionized magnesium by the University of Bremen (f).

\begin{figure*}
               \centering
              \includegraphics[width=\textwidth, scale=1]{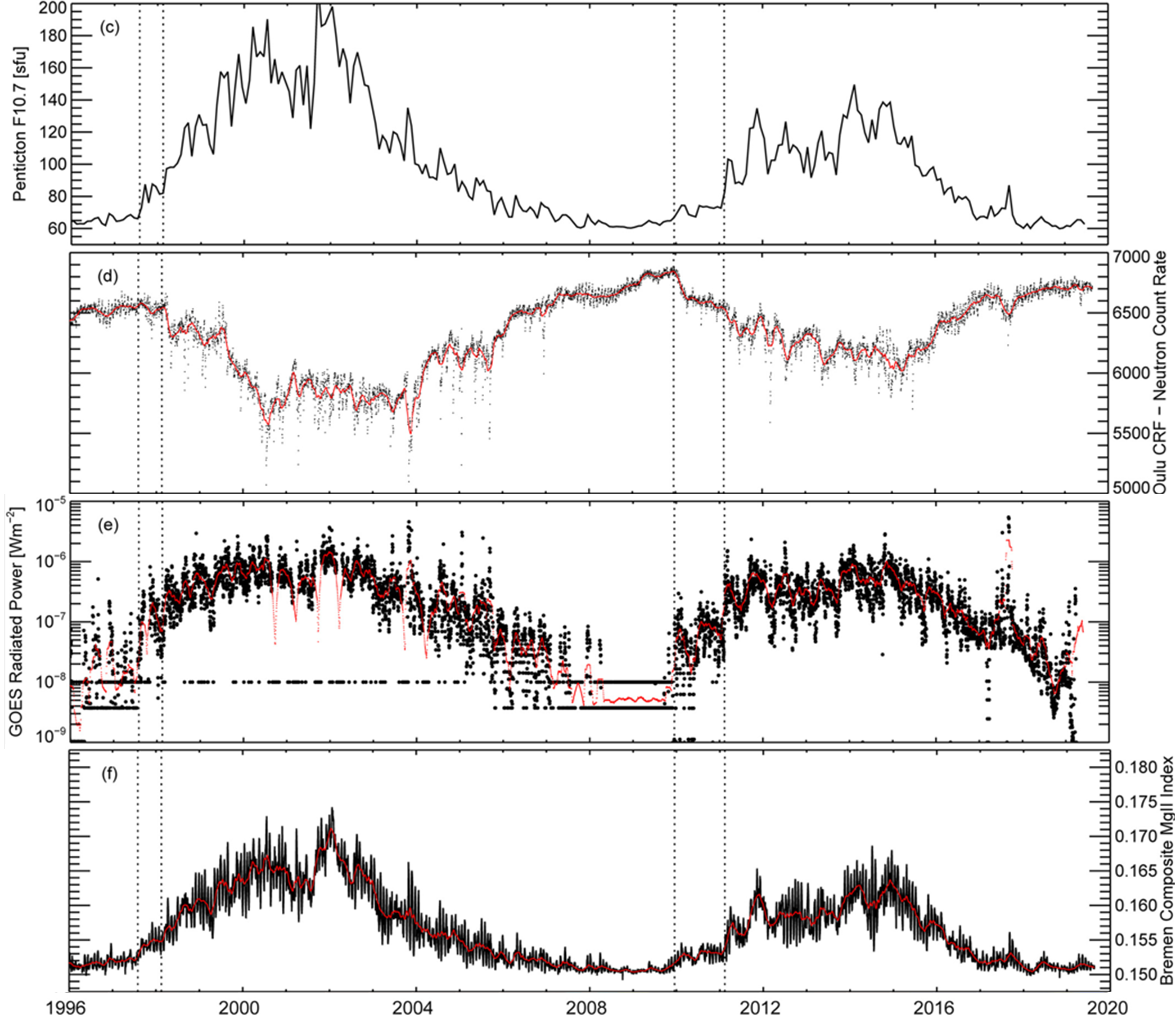}
                \caption{Solar activity from 1996 to 2020, observed by different instruments with various observables: c) F 10.7 A d) CRF e) X-Ray 1-8 A, f) UV emission. The vertical dashed lines represent the time-weighted ranges of the beginning of a new solar minimum }
                \label{fig:1:pectinton_cosmic}
                \cite{Terminator_Corr_Earth}
            \end{figure*}

On the other hand, extensive work has been conducted on Solar-Terrestrial connections, focusing on magnetic interactions that lead to changes in the Earth's global atmosphere. For example, in \citep{newell2002UV_solar_wind}, it is shown how the amplitudes of seasonal solar wind speed curves vary at the terrestrial heliospheric latitude according to the 11-year solar activity period. Some satellites in low Earth orbits are decelerated by atmospheric resistance, which depends on the atmosphere's density. Results in this regard have indicated that changes in ionization states caused by solar UV radiation produce modulations in the course of the solar cycle in the neutral density of the Earth's atmosphere \citep{hasting1996spacecraft}. \\

There is also experimental evidence that variations in global temperature are correlated with several spatial physical parameters: the long-term temperature anomaly follows solar radiation filtered by the cycle duration, which is a measure of solar cycle intensity \citep{pulkkinen2007space_weather}. Additionally, the global terrestrial cloud cover is correlated with the mean values of galactic cosmic ray fluxes. These are modulated by the solar cycle as solar wind pressure increases or decreases at its maxima and minima, respectively \citep{marsh2000low_clouds_cosmic}.\\

The exploration of this works concludes this section, introducing in an open manner the research intent being documented, albeit with a somewhat different exploration into the Sun-planetary atmosphere coupling, as will be evident in the subsequent development of this document.

\section{Dataset and method}

\subsection{Orbiter Mars Express and SPICAM-IR chanel}

On June 2, 2003, from Baikonur (Kazakhstan), aboard a Soyuz-Fregat launch vehicle, the Mars Express probe was launched towards the planet Mars. This marked the first long-term scientific mission carried out by the European Space Agency's scientific program, Horizons 2000. The mission comprised a three-axis stabilized orbiter with a high-gain antenna and various detection devices, along with the Beagle$-$2 lander module. The objectives of the mission were to conduct in-situ probing of Mars' subsurface, surface, and atmosphere. Starting from December 25, 2003, the orbiter's experiments initiated the acquisition of scientific data from Mars through an elliptical polar orbit \citep{Mex_begin}. The orbit of the Mars Express orbiter, designated as eq100, had, at the time of the last correction, a major semi-axis (a) of approximately 10,530 km, an inclination (i) of approximately 86.583 degrees, and an eccentricity (e) of approximately 0.943. It is evident from the orbiter's inclination that it orbits Mars with a quasi-polar trajectory, and with an orbital period of approximately 6.645 hours. 

Among all the scientific experiment instruments equipped on the Mars Express orbiter, the SPICAM (Spectroscopy for the Investigation of Characteristics of the Atmosphere of Mars) is the one responsible for atmospheric sounding through stellar occultation techniques. The data collected consists of spectroscopic images obtained in two distinct channels, one in the UV range and the other in the IR range. The UV channel senses wavelengths in the range of 118-320 $\nu$m, while the IR channel covers the range of 1-1.7 $\mu$m. The spectral resolution of the IR channel of SPICAM varies between 0.4-1 $\times$ 10$^{-3}$ $\mu$m. The spectral resolution power $(\lambda/\Delta \lambda)$ of this channel is approximately 1800 at 1.6 $\mu$m and approximately 2400 at 1.1 $\mu$m \citep{fedorova2018water}.

SPICAM is an extremely lightweight spectrometer, with a mass of 0.8 kg. It operates based on the principle of acousto-optic light filtration. In the IR channel, SPICAM possesses a sufficiently high spectral resolution to discern absorption features of H$_2$O in sunlight reflected by the planet. The instrument's field of view for nadir observations is 1$^\circ$, which corresponds to 4.5 km from the pericenter of the Mars Express orbit. The AOTF (Acousto-Optic Tunable Filter) spectrometer is a collaborative effort involving three institutions: the Space Research Institute (IKI) of Moscow, the Aeronautics Service of CNRS in France, and the Belgian Institute for Space Aeronomy. The spectrometer is developed and constructed in Russia, integrated into SPICAM, and calibrated in France. Certain mechanical parts of SPICAM-IR and one of the solar loading ports were manufactured in Belgium \cite{korablev2006spicam}. The AOTF  incorporated in SPICAM is the first device used for deep-space sounding, and comprehensive details of its extensive applications can be found in abundant literature on the subject. For instance, readers can refer to calibration specifics and the adapted method in \cite{AOTFgeorgiev2002, AOTF_moon, AOTF1996model}. The AOTF is optically tuned and operates based on the Bragg diffraction principle of an incident beam onto the ultrasonic acoustic wave excited within a birefringent crystal. The birefringent crystal possesses the property of splitting a light beam into two differently polarized components. Additionally, there is a built-in piezoelectric transducer designed for converting electrical signals into mechanical vibrations. When a radiofrequency (RF) signal is applied to the piezoelectric transducer, it generates mechanical vibrations in the birefringent crystal. These vibrations create an acoustic wave within the crystal, which interacts with the incident light beam. The Bragg diffraction principle comes into play when the acoustic wave inside the crystal encounters the light beam. The acoustic wave acts as a periodic grating that alters the propagation direction of the light beam, causing its diffraction. Consequently, new wavelengths corresponding to diffraction maxima are generated. By adjusting the frequency of the RF signal applied to the piezoelectric transducer, the position and width of the transmitted wavelength band can be controlled. This way, the AOTF filter enables the selection of a specific band of light and blocks unwanted wavelengths. The integration time per unit spectral point is 5.6 ms with an AOTF radio frequency power of 1504. In this channel, recording an equivalent spectrum of 3984 spectral points takes around 24 s. During this time, the Line of Sight (LOS) of the instrument spans about 150 km on the surface of Mars. During one orbit of the probe, covering latitudes between 50$^\circ$–100$^\circ$, between 40 and 70 spectra are recorded. The spectral response function of the MEx (SPICAM) spectrometer is determined by the properties of the AOTF filter. The width and shape of this function are important since one of the main observables (the flexural vibrational mode band of H$_2$O) of the spectrometer is not fully resolved in the obtained spectra, in addition to the uncertainty associated with the continuum \citep{korablev2006spicam}. The spectral response function was measured during pre-launch calibrations using a spectral line source. The final radiometric calibration of SPICAM was achieved in flight by comparing it with simultaneous measurements from the OMEGA mapping spectrometer \citep{altieri2009_calibration_Omega}. The spectrometer sensitivity, defined by the Noise Equivalent Brightness (NEB), is approximately 0.12–0.15 W/m$^2$/sr/$\mu$m in the range of 1.3–1.4 $\mu$m \citep{korablev2006spicam}.\

\subsection{Context of Data reduction Level 4}

The concentration profiles of a chemical species at various altitudes in the atmosphere of Mars, or in general, any planet, result from an elaborate procedure that commences with capturing images of transmission spectra through the stellar occultation method. This method involves capturing spectra of light transmitted by a star when it becomes obscured in the LOS of an observer located at a point. The LOS is ideally a tangent line to the planet's surface, encompassing a cross-section of the atmosphere and the occulted star, the transmission spectrum of which is being taken. The data processed in the case of this investigation consists of solar occultation spectra obtained in the IR channel through the SPICAM instrument aboard the Mars Express orbiter. However, the derivation of H$_2$O concentration profiles at level 4 was not carried out.

The characteristic spectra contributed by the atmosphere's composition are obtained by dividing all spectra acquired during an occultation by a solar (or stellar) reference spectrum observed outside the atmosphere. In the IR range of the SPICAM camera, information about atmospheric layers below 20 km cannot be extracted \citep{korablev2006spicam}, as the opacity of the Martian haze is so high that SPICAM struggles to detect reliable absorptions below this atmospheric layer. In the work by \cite{maltagliati2013occultations}, it is mentioned that the current confidence limit of the data, due to atmospheric extinction, fluctuates between 7 km when the atmosphere is very clear and 60 km when it is laden with dust. The time taken by the instrument to collect spectral points for solar occultations is 4 s, with observations at a full spectral resolution in a window of 609 spectral points between 1.341 and 1.469 $\mu$m. This window includes both the H$_2$O band at approximately 1.38 $\mu$m and the CO$_2$ band at approximately 1.43 $\mu$m \citep{maltagliati2013occultations}.

When the AOTF captures transmission spectra images centered at different frequency bands, they are recorded in a database for subsequent processing and analysis. Due to the principle of solar occultation, the transmission spectrum is obtained directly from relative measurements, eliminating the need for photometric calibration of the instrument. This results in atmospheric transmission spectra based on the relationship between the spectrum at different points in the atmosphere and the solar spectrum collected at an altitude above 120 km from the surface of Mars, for any season and location. The reference solar spectrum for occultations is obtained by averaging spectra within height ranges between 120 and 170 km \citep{fedorova2021_H2O}.

Since each spectral point in an occultation spectrum is measured sequentially, each point corresponds to a different altitude. These points lie within a continuum, which is estimated using linear interpolation method between the absorption lines of the region of interest \citep{maltagliati2013occultations}. Once the continuum is identified in each spectrum, spectral noise is reduced using the Savitzky-Golay smoothing filter \citep{fedorova2021_H2O}. This filter is effective in reducing noise in a signal due to its local polynomial minimization approach, which captures and preserves the underlying signal structure while removing noise in the spectrum. Generally, this filter provides precise smoothing and can be useful in various applications where noise reduction is needed without losing relevant information. Readers interested in the technical aspect can delve into the original work of the creators of this mathematical filter in \cite{savitzky1964_filtro}.  This was calculated using the European Mars General Circulation Model (EMGCM). The data from this model are progressively stored in the European Mars Climate Database (EMCD).\\

Following the smoothing process, the H$_2$O density profiles are retrieved by contrasting the spectral profiles from SPICAM with those obtained in simulations by varying parameters within the radiative transfer model (European Mars General Circulation Model, EMGCM). The fitting of synthetic spectra to the data is accomplished through the iterative Levenberg-Marquardt algorithm \citep{fedorova2021_H2O}, which involves a damped least-squares fit. The term 'damped' is used because in its implementation, a parameter called the 'damping factor' is chosen for the model curve. This factor controls the amount of correction applied in each iteration of the algorithm. In each iteration, the Jacobian matrix of the objective function must be computed, containing partial derivatives of this function with respect to each model parameter. This step estimates the sensitivity of the objective function to changes in the model parameters. The stability of the fitting method is achieved when the model, through gradient descent, finds the global minimum that represents the closest approach of the model curve to the data-associated function, minimizing the error between the two curves. Gradient descent occurs by varying the parameters of the Jacobian matrix and the damping factor, initiating the iteration process of the Levenberg-Marquardt method. The adjustment of these parameters must reach a point of stability between the data and the calculated model curve for the recovery of H$_2$O and CO$_2$ concentration profiles. The stability achieved by the method is tested using Tikhonov regularization to prevent overfitting of the model onto the data \citep{fedorova2021_H2O}. Readers interested in delving into details and examples of using this method can refer to \cite{levenberg1944method} and \cite{marquardt1963method}. Finally, uncertainties in the density profile estimations are obtained through the covariance matrix of solution errors in the Levenberg-Marquardt method \citep{fedorova2021_H2O}.\\
All the aforementioned processing results in H$_2$O concentrations at the estimated detection limit of 7-9$\times$10$^9$ molecules/cm$^3$, and a range of concentration profiles spanning between 5$\times$10$^9$ and 10$^{13}$ molecules/cm$^3$.

\subsection{Extension method from calibrated data of SPICAM-IR (Level 4)}

The data processing of SPICAM from Mars Express, following the procedure specified in \cite{fedorova2021_H2O} and \cite{maltagliati2013occultations}, generates lists of numbers representing physical quantities. These data can be freely downloaded through the \href{https://ode.rsl.wustl.edu/mars/productsearch}{\textbf{Mars Orbital Data Explorer}} linked to the Planetary Data System (PDS) Geosciences Node at NASA. The results are values derived from spectroscopic measurements of the vibrational states of water molecules present at different altitudes in the Martian atmosphere. These are well-packaged lists in three data columns: the first corresponds to surface heights, the second to the measurements of concentrations of different chemical species (H$_2$O for the IR channel and CO$_2$ for the UV channel), and the third to the uncertainties in the estimates of these chemical species. These three columns are stored in separate files for each date according to SPICAM's orbital detection, packaged in .TAB files with the NASA-designated format. For example ORBXXXXXAY\_H2O\_DENSITY.TAB, where the five-digit values $"$XXXXX$"$ correspond to the orbit number associated with the referenced dataset, $"$AY$"$ represents the orbit sequence number indicating the order in which the data was collected for orbit $"$XXXXX$"$ (A1, A2,...), and the letters correspond to the type of data being referred to. In our case, it's water vapor density: H2O\_DENSITY.\\

Each data package ".TAB" has a list of labels characterized by the same previous name, except that the file format corresponds to ".LBL". These labels contain detailed information about the derivations of the calibration level datasets in question, i.e., Level 4. In this format, you can see all the parameters associated with the ".TAB" data package. Notable among these are: The start date of the stellar occultation in the orbit number $"$XXXXX$"$, and the end date of the same. Intercept points of latitude and longitude for that dataset.
Solar longitude (position of the sun in the Martian sky). Data processing level identifier. Subsequently, under the encoding command in Vash /* DATA OBJECT DEFINITION */", you can see the labels for the three data columns that appear in the corresponding ".TAB" file, the numerical format of the data, and the physical units corresponding to each column. Thus, for altitudes, you can find 'Altitude of the line of sight above the Mars ellipsoid in km,' and for atmospheric H$_2$O concentrations and their respective errors: 'gas concentration in cm$^{-3}$.\\

\begin{figure*}
\centering
\includegraphics[width=1\textwidth]{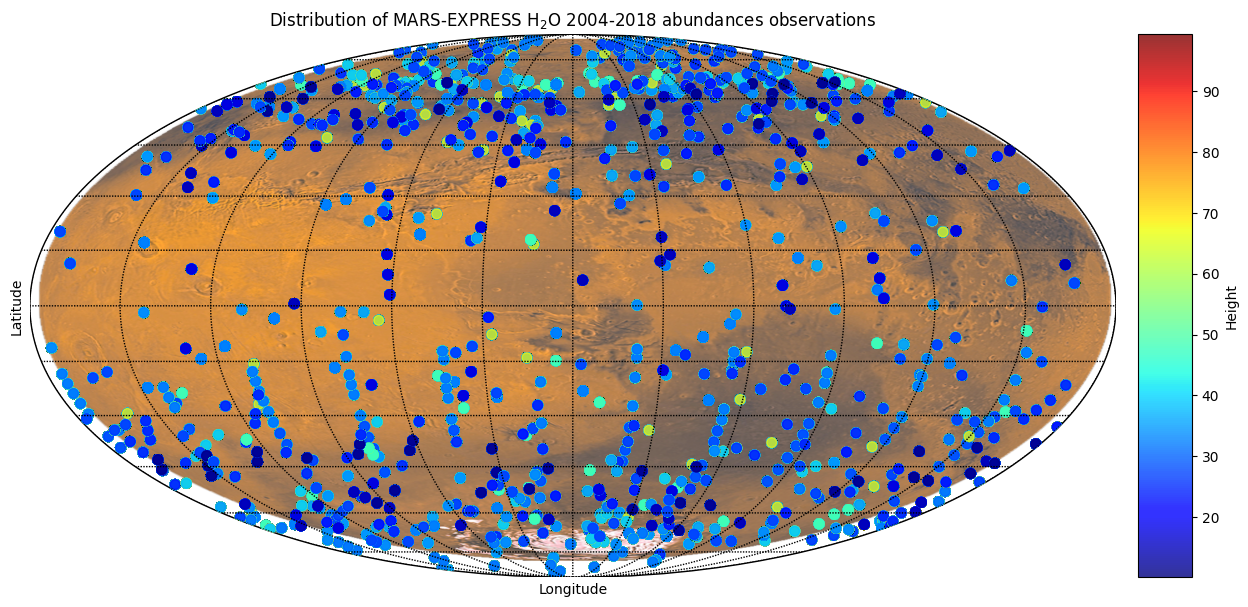}
\caption{Spatial distribution of atmospheric H$_2$O detections made by SPICAM of MEx.}
\label{fig:4:Marte_distrib}
\end{figure*}

The result of merging the two file formats described for the data volume (corresponding to more than 3,000 pairs of files) is a data table, which can be accessed in the "Data\_Derived\_ME\_H2O.xlsx" item on the Google Drive repository \href{https://drive.google.com/drive/folders/19-KLim7x3I1UOZyGYyvqF57r5XSeulEW?usp=sharing}{\textbf{Mars Atmosphere Analysis}}. These data correspond to observations made between the years 2004 and 2018, spanning the entire surface of Mars and ranging in altitude from 20 to 100 km above the Martian surface. Figure \ref{fig:4:Marte_distrib} displays the spatial distribution of the dataset.

Although the code implemented for data analysis can be adapted to any range of latitudes and longitudes, there are statistical (data abundance) and physical criteria that support the selection of certain regions on the Martian surface for study. In the spatial distribution of the data in Figure \ref{fig:4:Marte_distrib}, you can see the regions most suitable for study at different altitude intervals. For example, there is an abundance of data near the poles, in contrast to equatorial regions, within a range of 30$^\circ$S to 30$^\circ$N. As expected, the regions selected for this study encompass latitudes with a sufficient abundance of data to shed light on the variability of the chemical species of interest (atmospheric H$_2$O). 
The delimitation of latitude angular ranges is also determined by data abundance (through visual inspection). Furthermore, the region is chosen considering the influence of Mars' seasonal variation. Remember that Mars' ecliptic obliquity is 25.19$^\circ$. For the specific purpose of this study, regions within the Martian tropics (between 25.19$^\circ$ S and 25.19$^\circ$ N) are excluded. The spatial region defined on Mars for our study is between 60$^\circ$-80$^\circ$ latitude N and 0$^\circ$-80$^\circ$ longitude \footnote{Although all the analysis conducted has longitude boundaries of 0$^\circ$-80$^\circ$, it can be applied to any longitude range less than 180$^\circ$} (with the idea of studying changes in one hemisphere of the planet), yielding results similar to those outlined in this research.

The spatial distribution of atmospheric H$_2$O concentrations, for the sample corresponding to detections between the years 2004 and 2018, in the constrained region, is categorized into vertical profiles with altitude ranges of 10 km between 20 km and 100 km. This altitude interval is chosen in such a way that the variability in the concentration of the chemical species of interest is not due to changes in the vertical column due to pressure variations at different levels of the atmosphere, which would introduce errors into our analysis. In other words, the altitude range for a dataset to be studied must be below Mars' mean scale height, HS$_M$ = 11.1 km.

\begin{figure*} 
\centering
\includegraphics[width=1\textwidth]{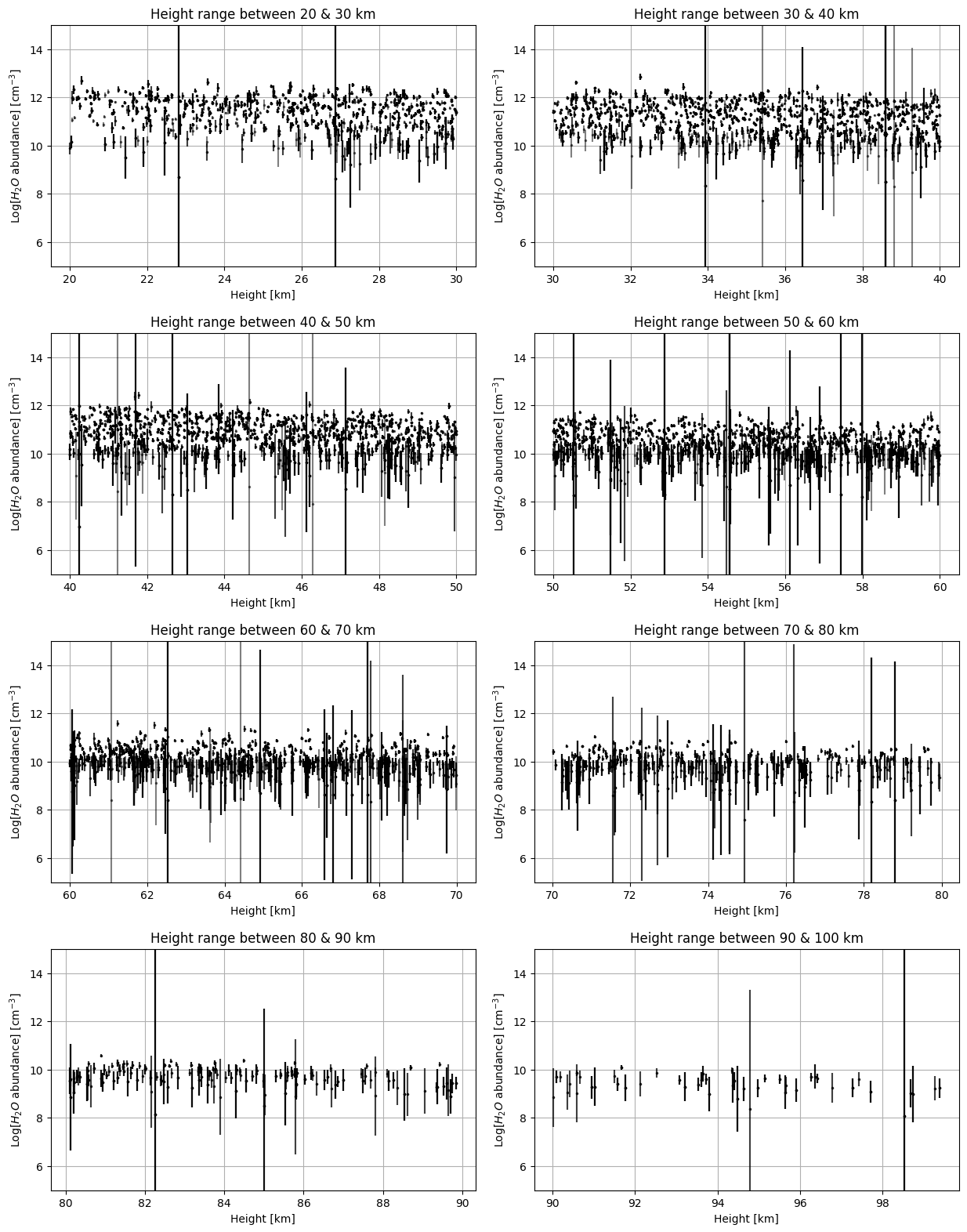}
\caption{Dispersions of H$_2$O concentrations for different altitude ranges. Note the different sizes of the error bars as a representation of the degree of confidence in the available detections.}
\label{fig:4:dispersion_all}
\end{figure*}

The concentration data for H$_2$O derived in the work of \cite{fedorova2021_H2O} come with associated uncertainties, represented in the data as error bars. To visualize the data, error bars are obtained based on the data distribution under a logarithmic fit. This process involves modifying the uncertainty bars by performing a corresponding error propagation analysis on the data uncertainties. The results can be seen in Figure \ref{fig:4:dispersion_all}. In this representation of the data, we can visualize how reliable the results are for subsequent analysis, in terms of the length of the error bars for different altitude ranges and the quantity of data available for sampling in each range.

Following the previous analysis, we represent the data within a time window from 2004 to 2018. H$_2$O concentrations are averaged per daily date and plotted in contrast to the 10.7 cm radio solar flux data, using an identical time window to ensure the same temporal resolution for the two signals being compared: atmospheric H$_2$O concentration and the characteristic 10.7 cm radio solar flux. This characteristic flux is normalized in units of solar flux, where 1 SFU = 10$^{-22}$ W m$^2$ Hz$^{-1}$. The daily solar flux data in the radio spectrum comes from the NRLMSISE-00 empirical data simulator. These are F 10.7 cm solar flux data measured daily by geostationary satellites in the GOES satellite network and serve as the main input parameter for the Earth's atmospheric simulator NRLMSISE-00. It also serves as the reference parameter for flux against which the solar activity signal is constructed.

\begin{figure*}
\centering
\includegraphics[width=0.9\textwidth]{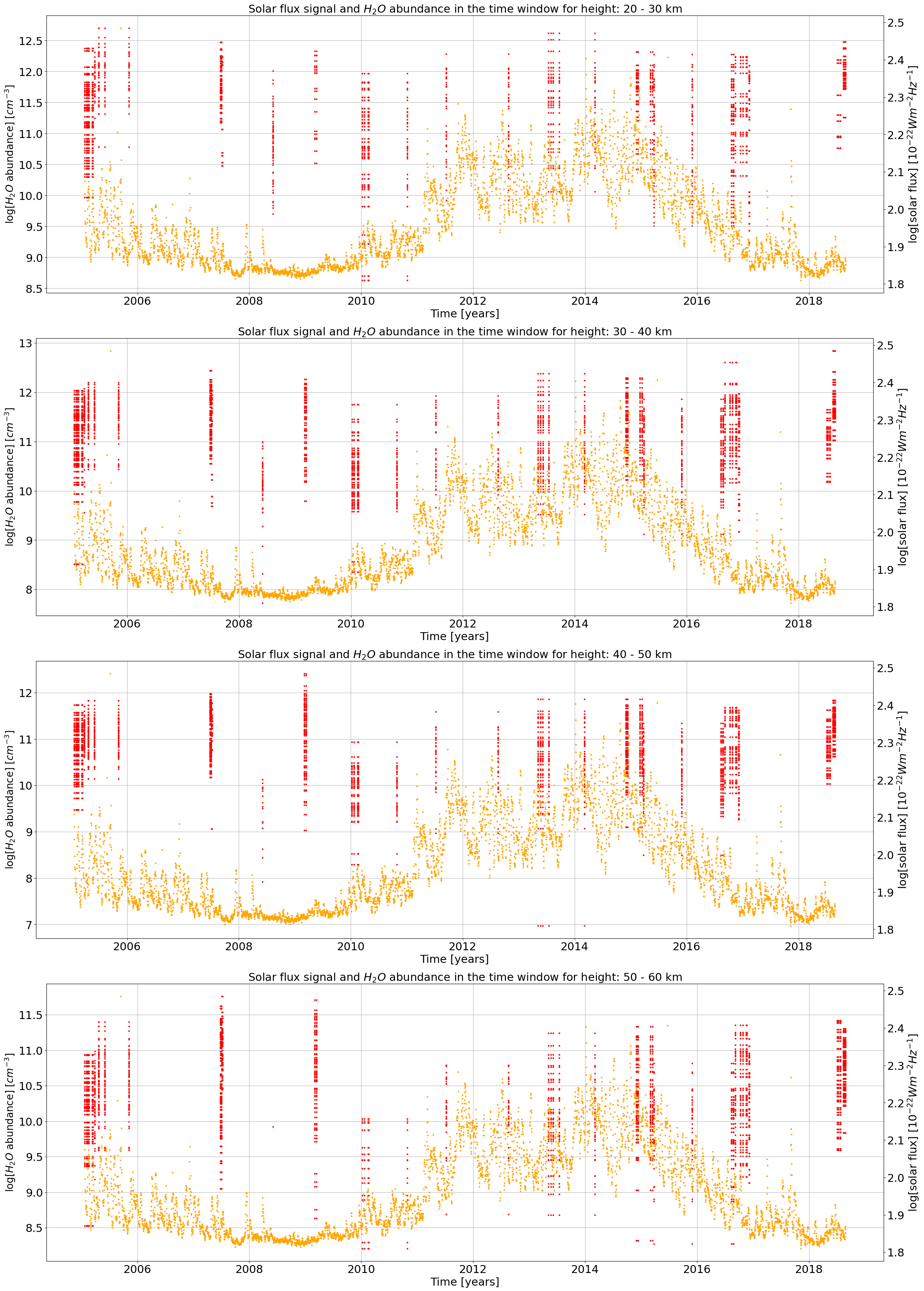}
\caption{Temporal dispersion of H$_2$O concentrations, for a range of heights 20-60 km. Note the homogeneity of the yellow solar flux signal in contrast to the loose H$_2$O concentration data.}
\label{fig:4:Mtemporal_h20-60}
\end{figure*}

\begin{figure*} 
\centering
\includegraphics[width=0.9\textwidth]{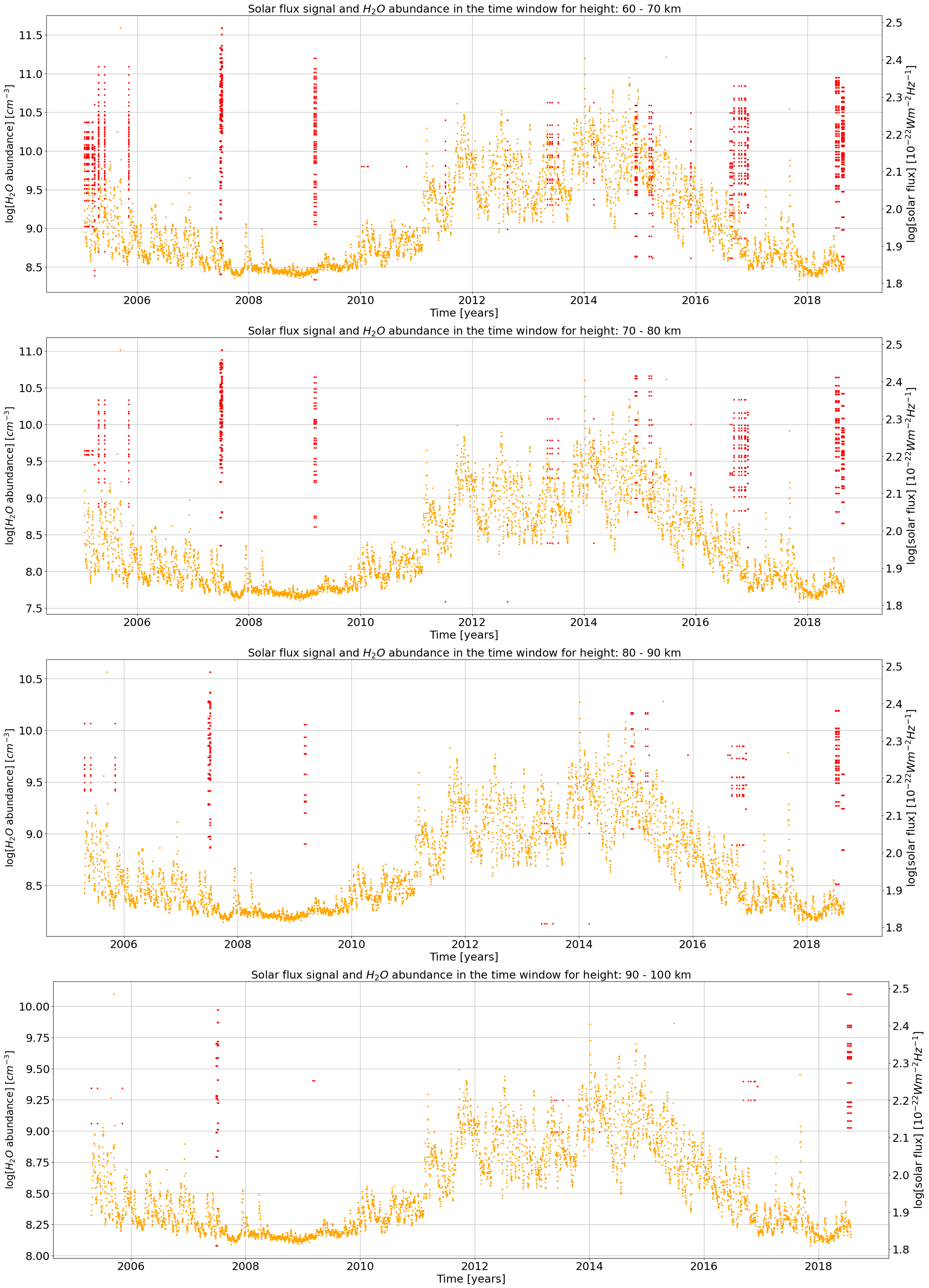}
\caption{Temporal dispersion of H$_2$O concentrations, for an altitude range of 60-100 km. Notice how the H$_2$O concentration data set decreases considerably for the last height ranges sampled.}
\label{fig:4:Mtemporal_h60-100}
\end{figure*}

Figures \ref{fig:4:Mtemporal_h20-60} and \ref{fig:4:Mtemporal_h60-100} illustrate that, for the characteristic region under study, there is no uniformly distributed dataset of atmospheric H$_2$O over time. Consequently, conducting an objective analysis of the relationship between the two signals (solar flux and H$_2$O concentration in the nth altitude range) is challenging. This is due to the sampling frequency of SPICAM and the fact that the SPICAM spectrometer aboard Mars Express orbits Mars rapidly, capturing between 40 and 70 spectra from different geographical points in the Martian atmosphere.\\

In summary, there are two signals available: solar flux at 10.7 cm radio, with uniform daily time sampling, and the second signal: data associated with atmospheric H$_2$O concentration at different altitudes, with non-uniform temporal data distributions. An appropriate approach to studying the possible relationship between these two signals requires establishing a method for continuing the analysis when a dataset is not uniformly sampled in a time series. One possible approach is to construct and analyze the frequency spectra of the two signals using the classical Fourier method, and then compare them, visualizing characteristic frequency peaks that are common to both signals. The following section introduces this method in detail, which allows for an objective analysis of how to relate signals with these characteristics while investigating Martian atmospheric activity induced by periodic variability in solar activity.

\subsection{Deciphering Patterns with the Lomb-Scargle Analysis Method for Datasets}

Given the above considerations, a classic periodogram is not the most suitable method for inferring the frequency spectra of the two signals. When data sampling is not evenly distributed in time, the typical method used by astronomers, e.g., for light curves of stars, is the calculation of the Lomb-Scargle Periodogram (LS-P) \citep{LSP-Vardesplast}. This method involves estimating the frequency spectrum of a signal by Fourier analysis of that signal, implemented computationally through the discrete Fourier transform (DFT) \citep{Lomb_1976, scargle_1982}. The difference from the traditional method is that it implements a least-squares regression model on harmonic functions that match the data distribution, whether it is homogeneously or inhomogeneously distributed within a time window \citep{LSP-Vardesplast}.
When data is not homogeneously distributed on a given time grid, there are too many harmonic functions superimposed on the data distribution, and as a result, many frequency peaks can be generated. Some of these peaks are associated with real periodic phenomena, while others are associated with fictitious oscillatory behaviors that appear in the data distributions, in addition to the typical noise of a signal, i.e., the Nyquist sampling frequency. The Lomb-Scargle method largely compensates for this problem by incorporating a confidence probability threshold into the frequency spectrum, known as the False Alarm Probability (FAP). The FAP involves statistically estimating the confidence of a peak's height compared to the false background peaks that appear in the periodogram. This property (the FAP) depends on both the number of observations and their signal-to-noise ratio (SNR). For small and low SNR values, spurious peaks in the background become comparable to the height of true peaks in the signal, so that the relevant information that a signal might bring is lost within the noise threshold. The ability to analytically define and quantify the relationship between peak height and significance is one of the main considerations that led to the standard form of the Lomb-Scargle Periodogram \citep{scargle_1982}.

In this context, the goal is to recreate the frequency spectra of pairs of signals of interest (solar flux at 10.7 cm and H$_2$O concentration at the nth altitude range) and calculate their corresponding periodograms, hoping to find the presence of a peak above certain noise probability thresholds (FAP) of 1\%, 5\%, and 10\% at the characteristic period of the 11-year solar cycle. The method incorporated for the periodogram estimations, following the previously explored theoretical framework, is summarized in the operation of the Python library LombScargle included in astropy. It is implemented in a way that the powers of the periodograms are normalized using the standard normalization method, where normalization is done through the data residuals around a constant reference model. Normalization allows the power peaks to be scaled within the range of 0 and 1 with dimensionless units, as well as the estimation of noise thresholds, determined through the FAP.

\subsection{Method testing with NRLMSISE$-$00 terrestrial datasets}

NRLMSISE$-$00 is an empirical model of the Earth's atmosphere that is fed with data packages every 2 months. It was developed in the year 2000, based on the previous models MSIS$-$86 and MSISE$-$90, by \cite{picone2002nrlmsise} with the purpose of making measurements of weather and climate for the knowledge of aerial and space exploration units. The estimates made by the NRLMSISE$-$00 model, using input data of solar flux at radio frequency F 10.7, produce numerical densities of He, O, O$_2$, N, N$_2$, Ar, H, H$_2$, pressure, and temperature estimates at different altitude levels relative to the Earth's surface. 

The NRLMSISE$-$00 model simulates the thermosphere through the Bates-Walker equations, which represent the basic profiles of temperature and density of chemical species as analytical functions of altitude. These equations are exact solutions of the atmospheric thermal and diffusive equilibrium model \citep{walker1965analytic}. Below an altitude dependent on the chemical species in a range of 160 to 450 km, the profiles differ from diffusive equilibrium in progressively larger amounts as altitude decreases, transitioning to a fully mixed state at a turbopause altitude of 100 km. At mesosphere altitudes, although there are data for mass density and temperature, the volume of data is too small. Therefore, the variations in the chemical species that the model reproduces mostly result from estimates made by equations modeling atmospheric behavior, such as the ideal gas laws \citep{hedin1991extension_NRLMSISE}. This holds true for altitudes ranging from 0 to 120 km. This approximation is valid for lower levels of the atmosphere, as it is modeled as a closed thermodynamic system within this altitude range.

\begin{figure*} 
\centering
\includegraphics[width=1\textwidth]{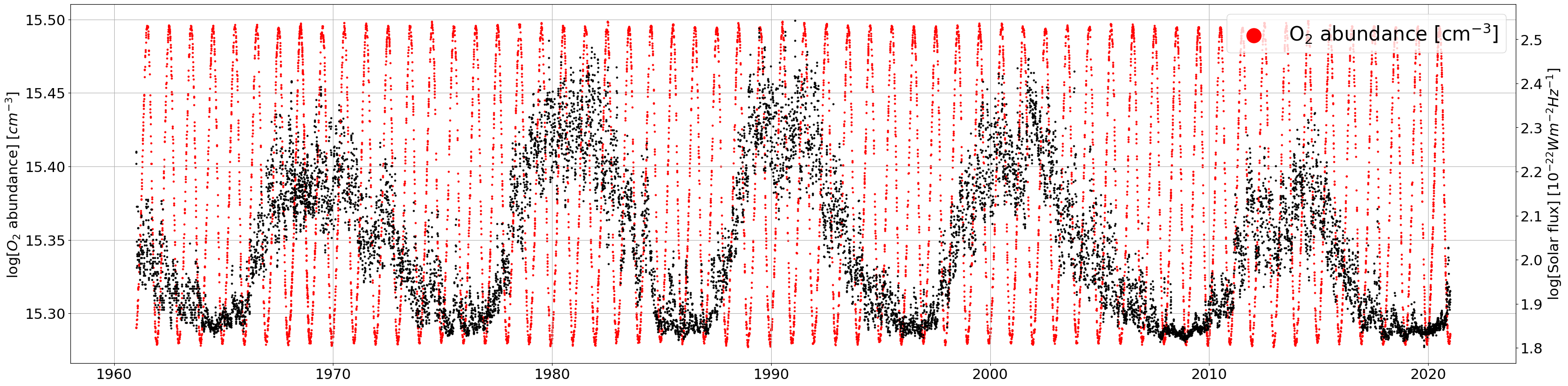}
\caption{Variability in the concentration of terrestrial O$_2$ at an altitude of 55 km obtained from the NRLMSISE-00 model. The oscillatory behavior of the signal is related to the Earth's orbital period. No oscillatory behavior that can be linked to the variable solar flux signal (in black) is observed in the signal.}
\label{fig:5:Earth_55_O2}
\end{figure*}

\begin{figure*} 
\centering
\includegraphics[width=1\textwidth]{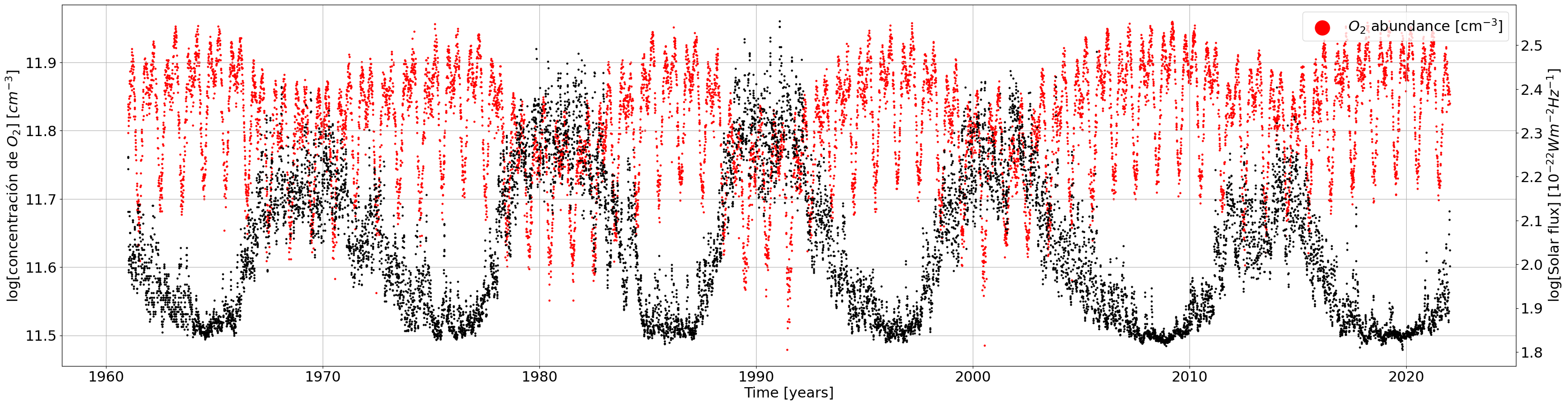}
\caption{Variability in the concentration of terrestrial O$_2$ at an altitude of 105 km obtained from the NRLMSISE-00 model. Oscillatory behavior associated with the Earth's orbital period can be observed in the O$_2$ signal, as well as oscillatory behavior associated with the solar cycle (in black), with which it appears to be anti-correlated.}
\label{fig:5:Earth_105_O2}
\end{figure*}

In this work, data from the empirical model NRLMSISE$-$00 were used to study the variability of different chemical species. This was done for two specific altitudes, one in the inner atmosphere at 55 km and the other in the lower ionosphere at 105 km altitude, both within a time window spanning from 1961 to 2021. Concentration estimates of the various chemical species were made based on daily averages for the same geographical coordinates on Earth, namely Latitude: 55°N, Longitude: 45°W. It's worth noting that an empirical parameter used as input in the model is the detection of the radio flux intensity at F 10.7 cm, which is recorded by Earth's orbiters such as GOES.
By employing the NRLMSISE$-$00 model data and representing it for the two aforementioned altitudes, for the chemical species O$_2$, we obtain Figures \ref{fig:5:Earth_55_O2} and \ref{fig:5:Earth_105_O2}. These figures depict the temporal distribution of simulated observations or detections over a 59-year time window, corresponding to 5.36 solar cycles. The magnitudes on the y-axis for each of the signals correspond to the concentration or density of the chemical species sampled in the simulation, in log[particles/cm$^3$]. The black line represents the solar flux intensity signal at 10.7 cm radio, also normalized logarithmically.

The reader may observe how the signal of O$_2$ concentration variation at an altitude of 105 km appears to be anti-correlated with the signal of variable solar flux activity as if they were two wave systems in a state of destructive superposition. Conversely, for the same chemical species at an altitude of 55 km, there doesn't appear to be a relationship between the two signals of interest. This approach to interpreting variable activity, both in solar flux and various chemical species, is significant as it lays the conceptual groundwork for subsequent treatment as signals in time windows with a composition of vibrational modes or frequencies, each associated with a specific oscillatory phenomenon within a particular time range. Consequently, these signals can be decomposed into frequency spectra through mathematical treatment using P-LS, as explored in the previous section of this chapter.

\begin{figure*} 
\centering
\includegraphics[width=1\textwidth]{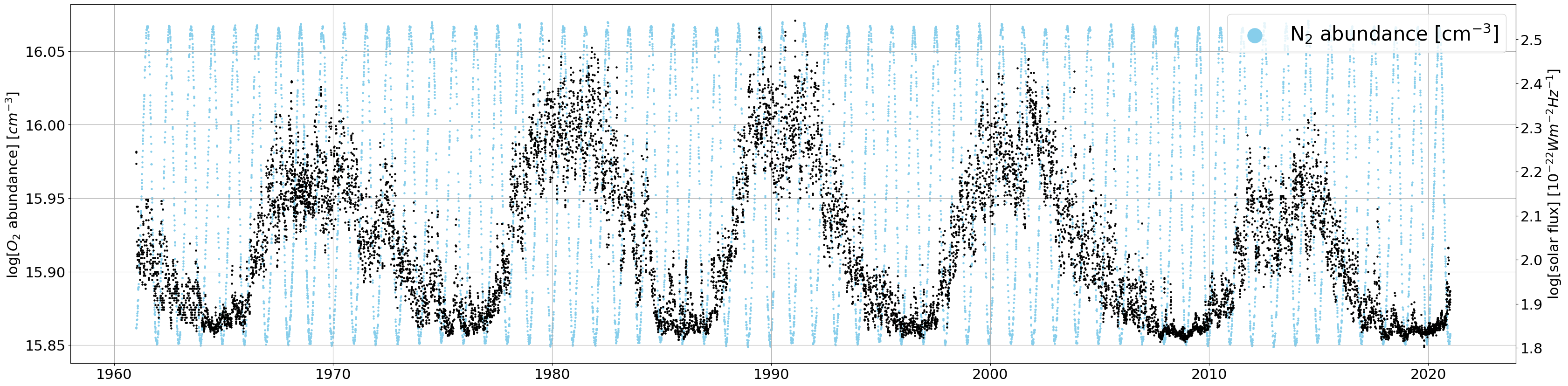}
\caption{Variability in the concentration of terrestrial N$_2$ at an altitude of 55 km obtained from the NRLMSISE-00 model. This signal exhibits similar behavior to the previous one; oscillations occur periodically with a frequency in sync with the Earth's orbital period.}
\label{fig:5:Earth_55_N2}
\end{figure*}

\begin{figure*} 
\centering
\includegraphics[width=1\textwidth]{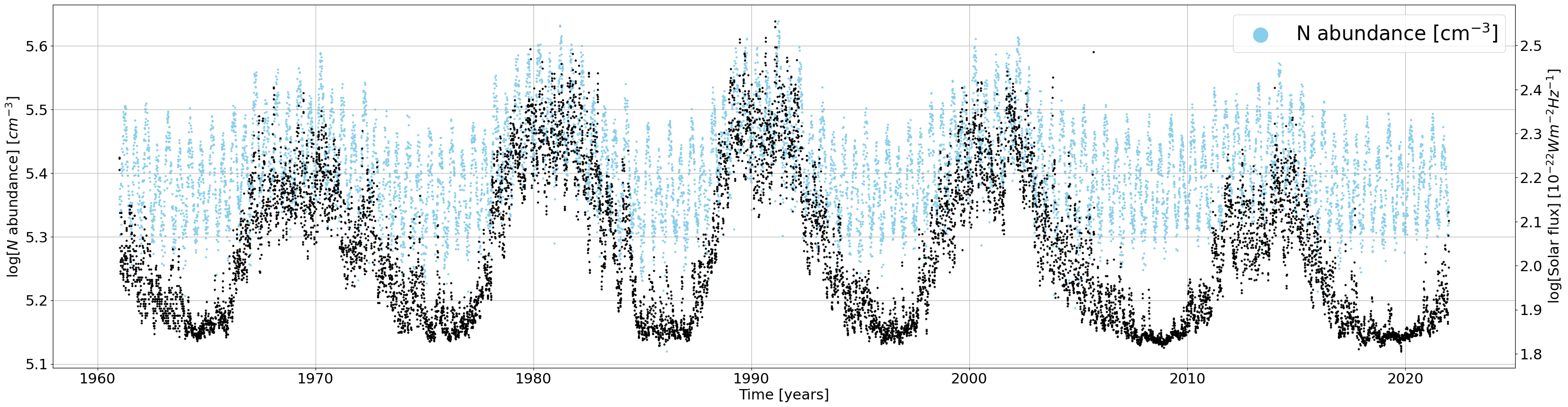}
\caption{Variability in the concentration of terrestrial N at an altitude of 105 km obtained from the NRLMSISE-00 model. The oscillatory behavior linked to the periodic activity of the solar cycle (in black) is highlighted, with which it appears to be correlated.}
\label{fig:5:Earth_105_N}
\end{figure*}

Figures \ref{fig:5:Earth_55_N2} and \ref{fig:5:Earth_105_N} display the signals of variability in the concentration of another chemical species (N$_2$) sampled for the test region at an altitude of 55 km and for N at an altitude of 105 km. In this case, contrary to the behavior of O$_2$, it appears that the two signals of interest are in phase and, therefore, to some extent, correlated. In general, the code and dataset for all analyzed chemical species can be found in the repository \href{https://github.com/NicolasLebrum/Planetary-Atmospheres.git}{Planetary atmospheres MEx}.

\subsection{The Periodogram Adapted to NRLMSISE$-$00 Data}

According to the previous exploration of the NRLMSISE-00 dataset, it became evident that in some cases, there seemed to be a pronounced correlation between the signals of variability in the concentration of certain chemical species at an altitude of 105 km and the solar cycle signal. This correlation will be further investigated by searching for the relationship between frequency peaks in the two signals using the LS-P method, as explored earlier.

\begin{figure*} [H]
\centering
\includegraphics[width=1\textwidth]{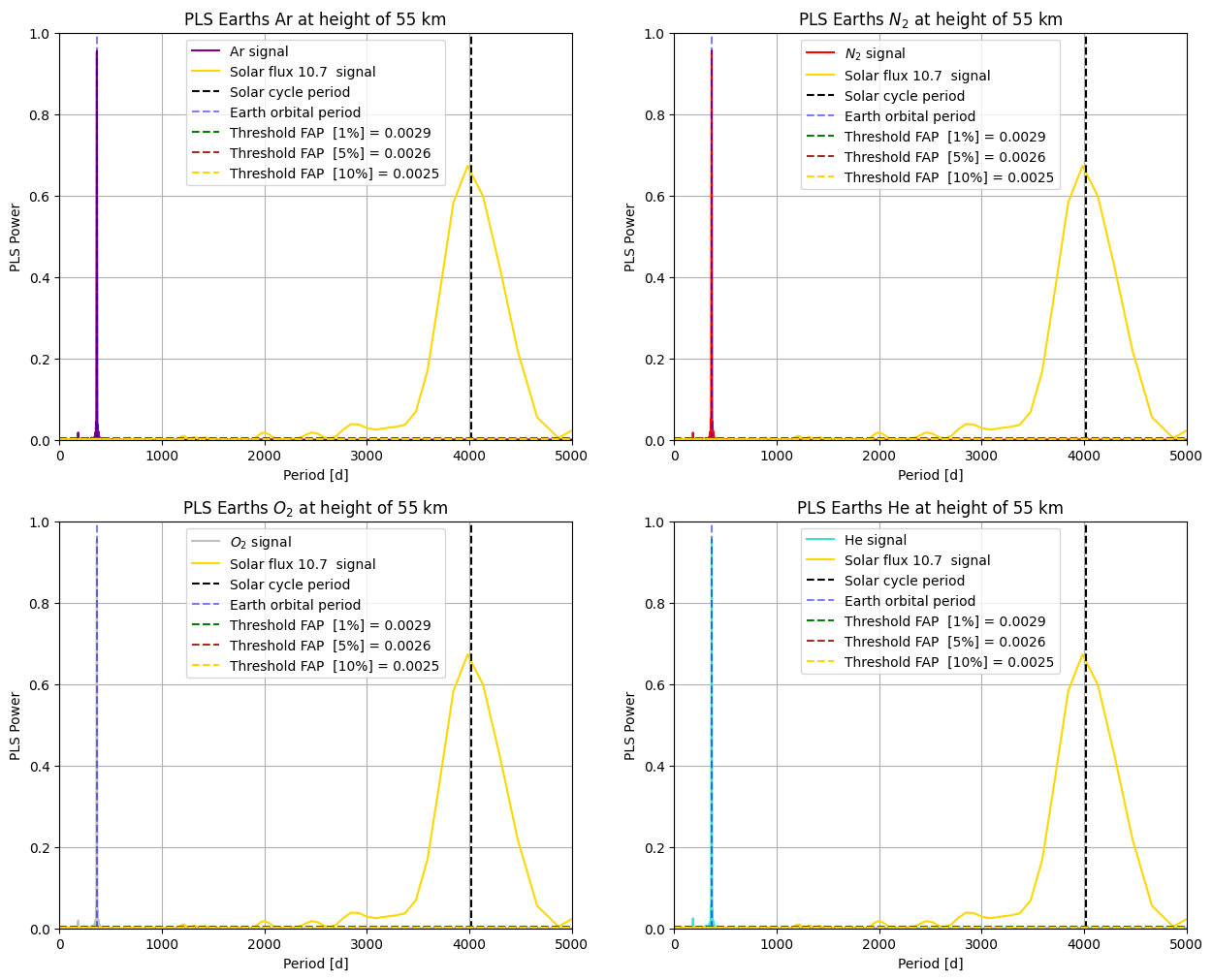}
\caption{LS periodograms obtained for the signals of the chemical species Ar, N$_2$, O$_2$, and He, contrasted with the periodogram of solar activity in 10.7 cm radio flux. Note that on the scale of the solar cycle period, no peak associated with each analyzed chemical signal is observed.}
\label{fig:5:Earth_55_LSP}
\end{figure*}

\begin{figure*} [H]
\centering
\includegraphics[width=1\textwidth]{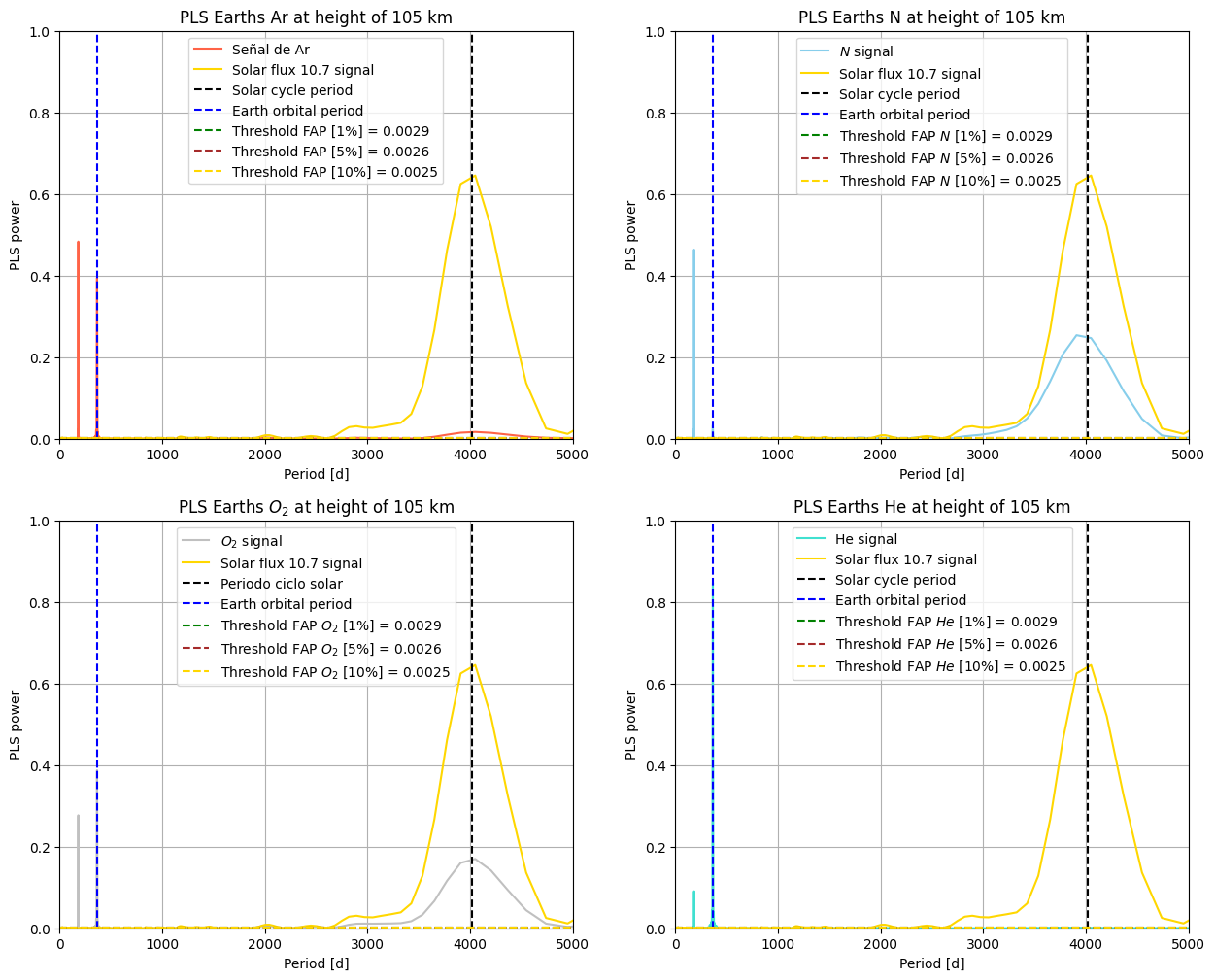}
\caption{LS periodograms obtained for the signals of the chemical species Ar, N, O$_2$, and He, contrasted with the periodogram of solar activity in 10.7 cm radio flux. Note that on the scale of the solar cycle period, the peaks of the two signals of interest align for each analyzed chemical species.}
\label{fig:5:Earth_105_LSP}
\end{figure*}

The performed periodograms have a maximum frequency threshold set at $\frac{1}{20 d} = 0.05 d^{-1}$. The choice of this threshold is based on maintaining a boundary where certain data continuity is not lost. Additionally, peaks associated with short-period phenomena are not of interest in this research. The peaks of interest are the Earth's orbital period frequency peak, $\frac{1}{365 d} \approx 0.00274 d^{-1}$, and the solar cycle frequency peak, $\frac{1}{4015 d} \approx 0.00025 d^{-1}$. Another criterion considered for the corresponding calculation of the periodograms is the number of neighboring points to a peak. This is particularly relevant when dealing with a signal constructed from non-uniformly spaced data in the time window to smooth the distribution of peaks in the calculated periodogram. In this regard, since the signals simulated through NRLMSISE-00 have daily periodicity, no adjustment was needed for this parameter to smooth the frequency spectra. The results for the different studied signals are shown in Figures \ref{fig:5:Earth_55_LSP} and \ref{fig:5:Earth_105_LSP}.

As expected, in accordance with the concentration data of different chemical species in the time window 1961-2021, there are prominent peaks in the spectral profiles of the periodograms, especially at an altitude of 105 km, in the lower part of the Earth's ionosphere. These periodograms reveal that the prominent peaks correspond to two characteristic periods. The first corresponds to half of the Earth's orbital period for all chemical species. This suggests that seasonal changes at this altitude (105 km) manifest as 6-month periods in the ionosphere more intensely than those present for the Earth's orbital period. The second period is related to the solar cycle period, where peaks are observed for all analyzed chemical species in the lower ionosphere at 105 km altitude. For the lower mesosphere at an altitude of 55 km, the most intense peaks in the periodograms correspond to the Earth's orbital period.

In connection with the above, it is important to consider that the lower mesosphere (between 50 and 80 km altitude) contains the smallest air mass (0.1\%) in proportion to the Earth's atmosphere, in addition to being the coldest region, reaching temperatures of -80$^{\circ}$C \citep{states2000thermal}. Due to these conditions, the air density is so low that it stimulates the production of turbulence and atmospheric waves (WG) that operate on large spatial and temporal scales \citep{she2022climatology_mesosphere_WG}. Finally, at this depth of the atmosphere, solar activity does not directly induce variable dynamics in the chemical concentration observables, as reflected in the solar cycle mark in the periodograms for this altitude.

The peaks present in the lower part of the ionosphere for all chemical species (except for He) align quite well with the peak associated with solar activity with an 11-year period. Note in Figure \ref{fig:5:Earth_105_LSP} that the periodograms of the two frequency spectra (chemical species signal and solar activity signal) have a peak that aligns quite precisely with the mark associated with the known 11-year solar cycle period. The precision in this case is due to the fact that the data used to calculate the periodogram are derived from a simulation that interpolates over data taken from the atmosphere using characteristic equations for modeling the Earth's atmosphere \citep{walker1965analytic, hedin1991extension_NRLMSISE}.

\section{Results of H$_2$O abundances in the Mars Atmosphere}

\subsection{Local analysis of atmospheric H$_2$O distribution in relation to solar activity }

To analyze the variation in the concentration of the atmospheric chemical species H$_2$O within the previously defined characteristic regions, especially while observing the range of variation of the specific chemical species in question, an average concentration of H$_2$O is assumed at a specific altitude within the range of altitudes for which data is available, with a concentration of 13.6$\times$10$^{12}$ particles/cm$^{3}$. This assumption is based on the calculation of the average H$_2$O concentrations within a height range with a characteristic data sample.\\

\begin{figure}
\includegraphics[width=\columnwidth, scale=1]{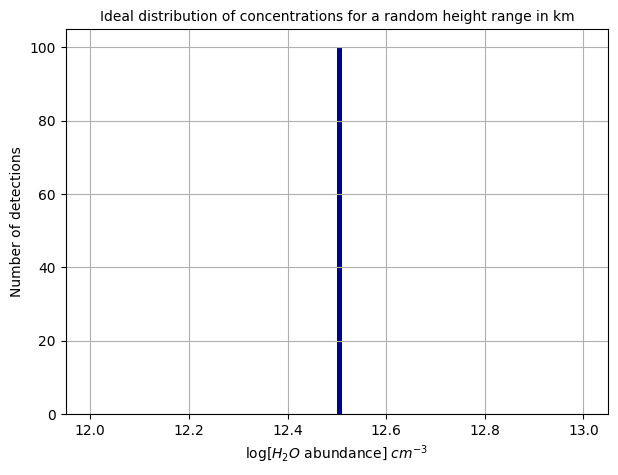}
\caption{Detection of idealized atmospheric H$_2$O concentrations on Mars: all have the same value.}
\label{fig:4:H2O_ideal}
\end{figure}

Taking as a reference an average concentration value of water vapor H$_2$O on Mars is an adaptable proposition to the fact that Mars is modeled as an ideal object with an atmosphere that is not influenced by the Sun throughout its orbit or by the planet's internal dynamics, but does contain water vapor in its composition. Under the previous assumption, any detection made regarding the concentration of H$_2$O will coincide with this average value (12.5$\times$10$^{12}$ particles/cm$^3$), as can be seen in Figure \ref{fig:4:H2O_ideal}.

\begin{figure*} 
\centering
\includegraphics[width=0.8\textwidth]{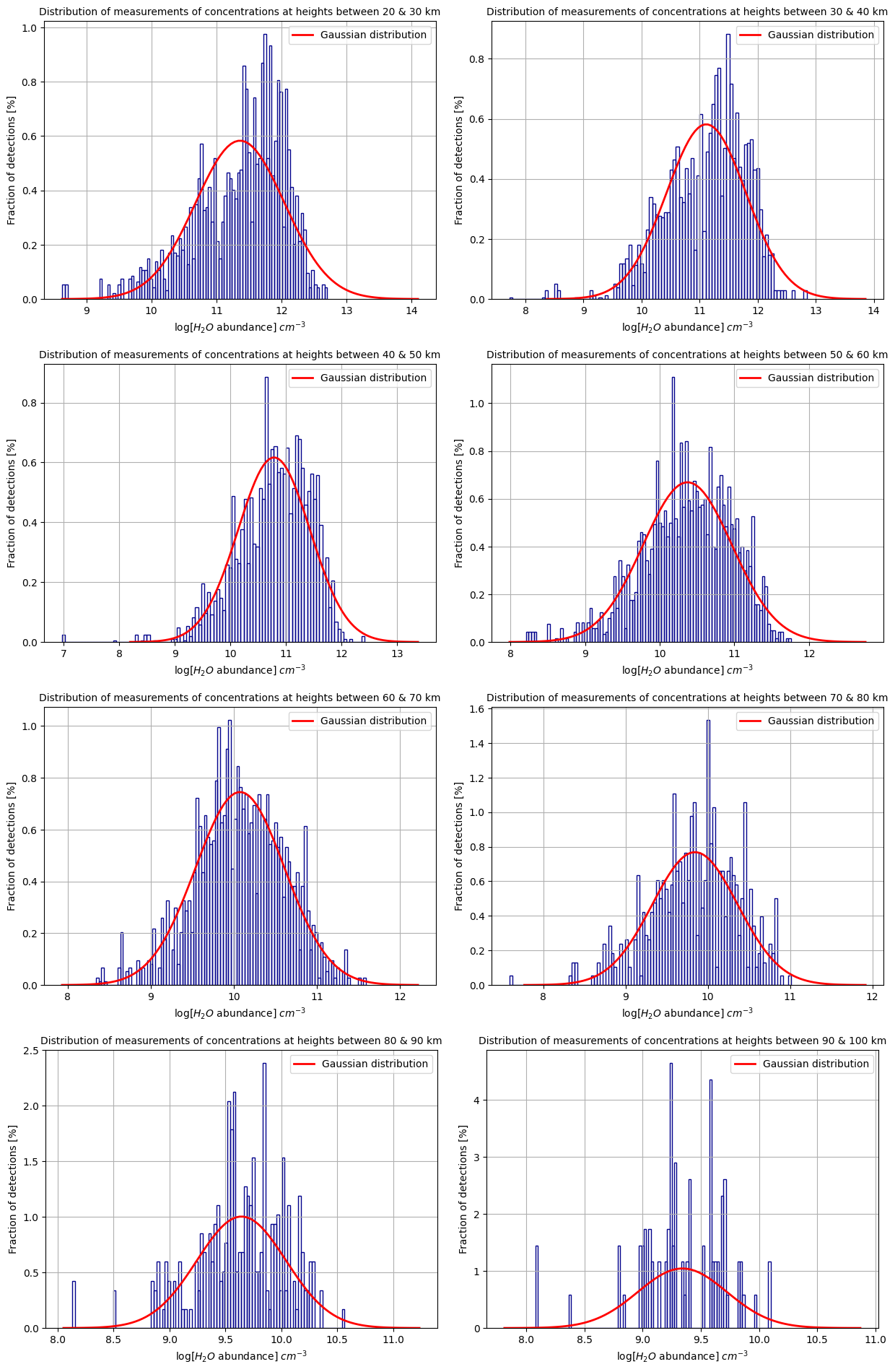}
\caption{Normalization of atmospheric H$_2$O concentration distributions for different altitude ranges, and Gaussian fitting of them. Note that for higher altitudes, the fitting seems not to hold.}
\label{fig:4:H2O_gauss_all}
\end{figure*}

The distribution of concentrations for different altitude ranges in the region of interest is illustrated in Figure \ref{fig:4:H2O_gauss_all}. These histograms (normalized) have a selection criterion of a number of classes (bins) corresponding to 100 so that 100 varieties in atmospheric H$_2$O concentration can be visualized. According to the statistical Sturgles' rule, with a sample that varies for each altitude range between $\approx$ 3850 and $\approx$ 200, for altitudes of 20-30 km and 40-50 km, respectively.

The dispersion of the data illustrated in Figure \ref{fig:4:H2O_gauss_all} suggests a Gaussian distribution regarding atmospheric H$_2$O detections, where the mean peak changes for each altitude interval in the sampled region. Indeed, in the normalized distributions illustrated in Figure \ref{fig:4:H2O_gauss_all}, it can be observed that in most altitude levels, the data appears to be distributed under Gaussian bells; however, for some altitude levels, the distribution of H$_2$O concentration detections deviates from being Gaussian. This occurs particularly for altitude ranges where our data sample is very small, around $\approx$200 data points for h = 90-100 km. 

\begin{figure}
\centering
\includegraphics[width=\columnwidth, scale=1]{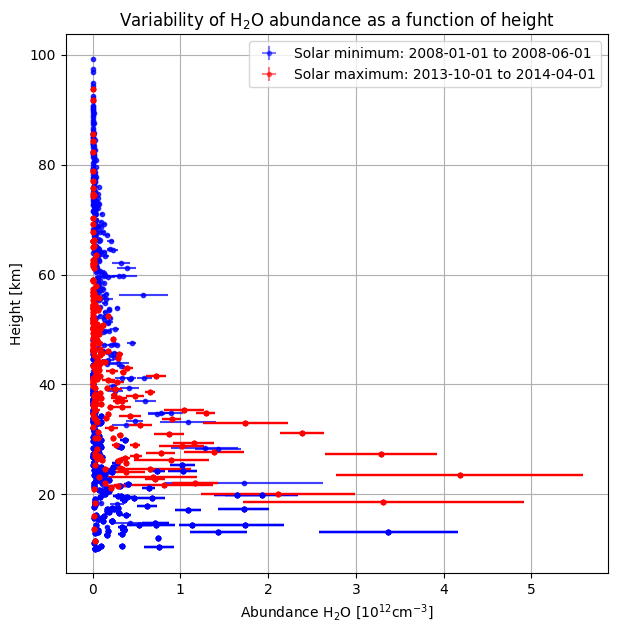}
\caption{H$_2$O concentrations during solar maximum and minimum in the altitude column between 20 and 100 km for the sampled region.}
\label{fig:4:H2O_solar_allh}
\end{figure}

After having a statistical characterization of the data sample derived in the work of \cite{fedorova2021_H2O}, which provides information about the range of variation in water vapor concentration, we proceed to analyze the expected variability bounds for atmospheric H$_2$O height profiles during solar minimum and maximum (events of particular interest in our study). These profiles are illustrated in Figure \ref{fig:4:H2O_solar_allh}. 

We observe that for higher altitudes, between 50 and 80 km, the distribution of H$_2$O concentrations within this range during a solar minimum is greater than during a solar maximum. This is expected since during a solar maximum, there is usually greater photospheric activity and, therefore, greater solar wind intensity, which is reflected in greater radio intensity according to the characteristic radio flux at 10.7 cm in our data. When there is higher solar activity, it is expected that the Martian atmosphere will heat up, and therefore, by the classic thermodynamic principle of thermal expansion, the concentration of atmospheric H$_2$O will be lower, in contrast to when the atmosphere cools during a solar minimum, where the concentration of atmospheric H$_2$O should increase.

Contrary to expectations, the above behavior is the opposite for altitudes below 50 km; during solar maximum, the concentration increases, and during solar minimum, it decreases. Therefore, the concentration distribution does not follow the expected behavior.

\begin{figure} 
\centering
\includegraphics[width=\columnwidth, scale = 1]{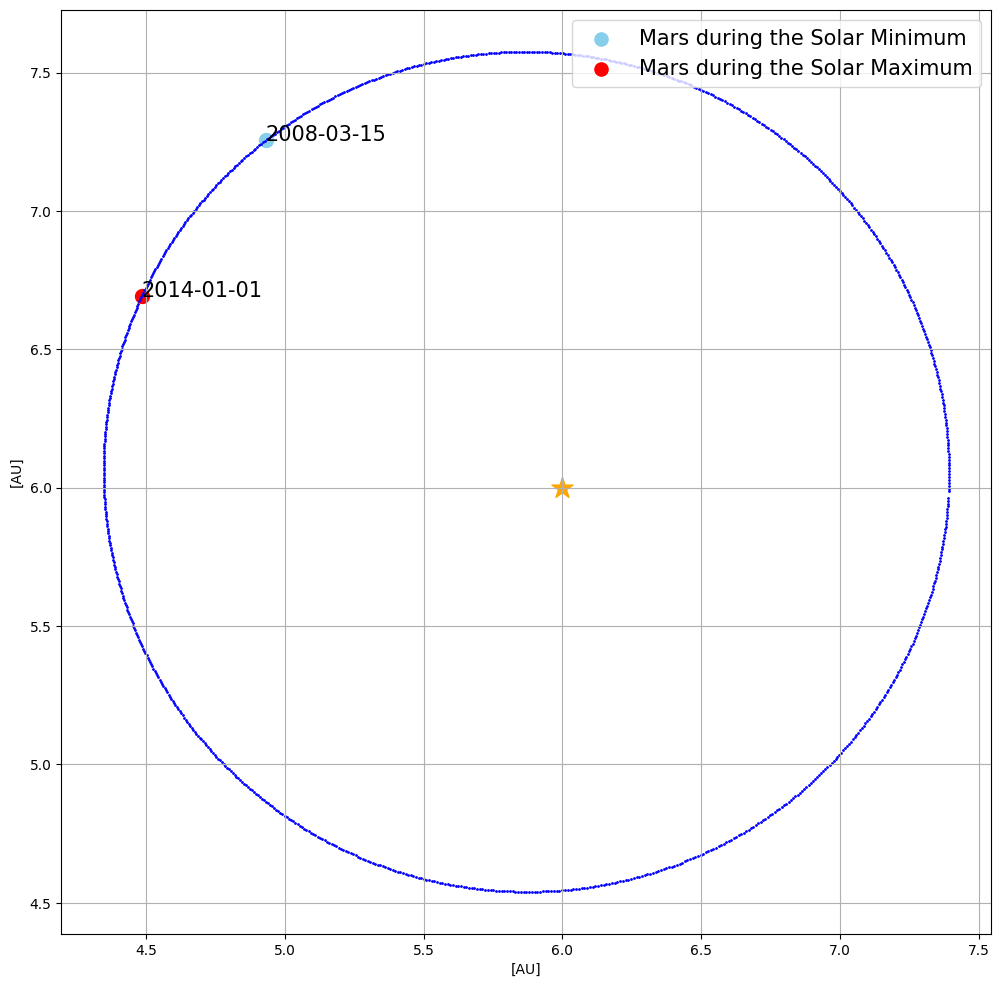}
\caption{Orbital position of Mars during solar maximum and minimum identified for the H$_2$O concentration data sequence. Note that Mars is close to the aphelion of its orbit in both situations. The Sun is represented by the yellow symbol.}
\label{fig:4:orbita_Marte_maxmin_solar}
\end{figure}

We ensure that the treated solar maximum and minimum correspond to a similar orbital position of Mars so that the data are influenced as little as possible by the seasonal behavior of Mars. This is done by observing the planetary ephemerides of Mars through NASA SPICE planetary geometry software \citep{acton1996ancillary,acton2018SPICE}, as shown in Figure \ref{fig:4:orbita_Marte_maxmin_solar}. This is directly related to the planetary system's geometry by calculating the r/a parameter, obtaining the Mars-Sun distance ratio at the dated moments as a reference for solar maximum and minimum. See Table \ref{tab:4:orbital_parameter}.

The results r/a solar minimum = 1.0821 and r/a solar maximum = 1.0932 reveal a discrepancy in orbital positions of 1.11$\%$, which translates into the reliability that the altitude profiles of H$_2$O concentration variability are due to solar activity and the planet's internal dynamics, which are not discriminated in our procedure and analysis.

\begin{table}
\centering
\begin{tabular}{|l|l|}
\hline
\textbf{Parameter (r/a)} & \textbf{Value} \\ \hline
Perihelion               & 0.9065         \\ \hline
Aphelion                 & 1.0931         \\ \hline
Solar maximum            & 1.0821         \\ \hline
Solar minimum            & 1.0932         \\ \hline
\end{tabular}
\caption{Parameter r/a for Mars' orbital positions of interest.}
\label{tab:4:orbital_parameter}
\end{table}

In addition to the above, in the work of \cite{GW_planetary}, the behavior of Mars' atmosphere during a solar minimum is studied to analyze Mars' atmospheric behavior due to its internal dynamics, minimizing the solar influence as much as possible. In their results, the authors describe how the atmospheric behavior in the high-altitude layers of Mars is influenced by wave-like effects produced in the intermediate and lower regions of Mars' atmosphere (GW). This could possibly be an effect that manifests in the change in H$_2$O concentration behavior below 50 km in altitude, deviating from the expected behavior.

\subsection{Coupling Mars' H$_2$O abundance variability with solar activity}

Previously, the LS$-$P method was introduced as a mathematical tool to assess the relationship between the periods of two signals. Data from the empirical NRLMSISE$-$00 model were used for both solar flux and variability in the concentration of different chemical species in the Earth's atmosphere at different altitudes. These data were employed as a preliminary step to test the mathematical LS$-$P method before directly analyzing the primary data of interest in this work, which is the variability in the concentration of H$_2$O on Mars.

The H$_2$O concentration data on Mars, derived from observations and detections by SPICAM on Mars Express, unlike the synthesized data from NRLMSISE$-$00, have associated uncertainties for each sampled altitude range. Consequently, an analysis of error and confidence can be established, as will be seen in the following sections. In Figures \ref{fig:4:Mtemporal_h20-60} and \ref{fig:4:Mtemporal_h60-100}, distributions in the time window 2004-2018 of H$_2$O concentration data at different altitudes were observed in contrast to the F 10.7 cm solar flux data. While it is challenging to discern any relationships in these temporal data distributions, unlike what can be done for data obtained from the NRLMSISE$-$00 model (see Figures \ref{fig:5:Earth_105_N}, \ref{fig:5:Earth_105_O2}, \ref{fig:5:Earth_55_N2}, \ref{fig:5:Earth_55_O2}), the incorporated LS$-$P method can shed light on this matter through the calculation of signal frequency spectra and their corresponding analysis. Frequencies related to possible systematic and correlated noise, such as the sampling frequency or Nyquist frequency, are well above the frequencies of interest, namely the frequency associated with Mars' orbital period and the frequency related to the solar cycle period. The search for the latter is established as the main objective of this research, and a correspondence between peaks of H$_2$O concentration variability for the highest altitude ranges and solar activity variability in F 10.7 cm radio is anticipated.

As detailed in the analysis corresponding to the NRLMSISE$-$00 model data processed using the LS$-$P method, the periodograms of Mars' signal have a maximum frequency threshold of 0.05 d$^{-1}$, corresponding to a period of 20 days. Additionally, an interpolation value for smoothing was established so that each peak in the periodogram had 15 points in its vicinity. This was chosen as it represents the minimum number of points at which the signal remains continuous without overfitting.

Applying the LS$-$P method to the H$_2$O concentration signals for the region of interest results in different periodograms, corresponding to altitude ranges of 10 km, between 20 and 100 km above the surface of Mars. This method is also applied to the solar radio flux data at 10.7 cm, derived from observations from the GOES satellite network within the same time window as the concentration signals, ensuring that the generated frequency spectra have the same temporal resolution. It should be noted that for any of the calculated periodograms corresponding to H$_2$O signals for all altitude ranges, there are three characteristic thresholds that determine the level at which frequency peaks or periods must exceed to be above the signal's noise level, manifested as the FAP. These thresholds correspond to the maximum levels of the peaks so that the processed signal has a 1\%, 5\%, and 10\% probability that the frequency spectrum does not correspond to a periodic behavior of the data but is instead a result of systematic noise in the processed signal. This systematic noise is usually associated with the sampling frequency and error correlation of each signal, as part of the original signal whose power spectrum is being calculated.

\begin{figure*}
\centering
\includegraphics[width=0.9\textwidth]{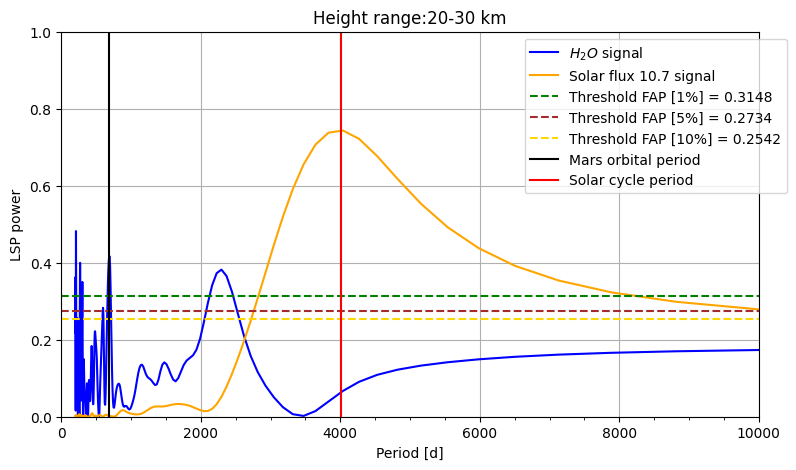}
\caption{Lomb-Scargle Periodogram for the altitude range 20-30 km.}
\label{fig:6:LSP_H2O_20-30}
\end{figure*}

In the periodogram corresponding to the altitude range 20-30 km (Figure \ref{fig:6:LSP_H2O_20-30}), the periodic peak of the signal aligned with Mars' orbital period of 687 days is highlighted, along with a peak that stands out above all false alarm probability thresholds at approximately 2250 days. This peak turns out to be a discovery in the data and is common to all altitude ranges analyzed. For this altitude range, there is no apparent peak associated with the solar activity period with a significant statistical relevance. In summary, there does not seem to be a relationship between atmospheric activity and variability in the concentration of H$_2$O on Mars for the 20-30 km altitude range.

\begin{figure*} 
\centering
\includegraphics[width=0.9\textwidth]{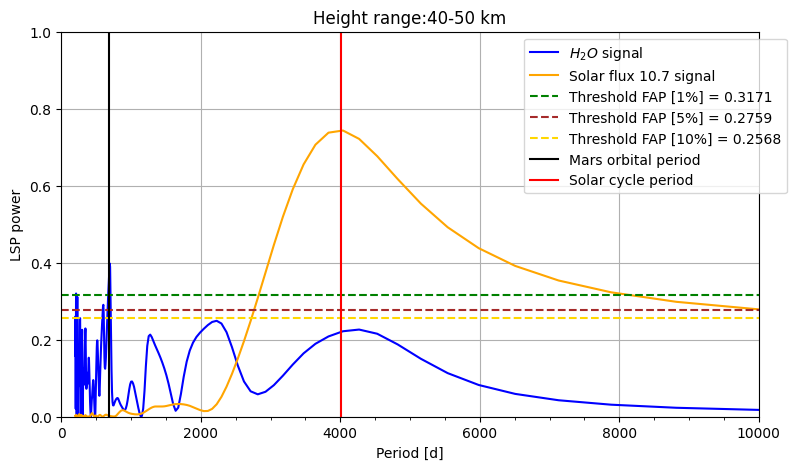}
\caption{Lomb-Scargle Periodogram for the altitude range 40-50 km.}
\label{fig:6:LSP_H2O_40-50}
\end{figure*}

The synthesis of the analysis above is very similar to that for the altitude ranges 30-40 km (see Figure \ref{fig:6:LSP_H2O_20-30}) and 40-50 km (see Figure \ref{fig:6:LSP_H2O_40-50}), except for the fact that the only peak with statistical significance according to the incorporated method (FAP) is the one corresponding to Mars' orbital period. It should be noted for both ranges that the peak pronounced at approximately 2250 days is present, with the difference from the first range (20-30 km) being that this peak does not exceed any of the corresponding false alarm probability levels. For the 40-50 km altitude range, a small peak at the solar cycle period is defined but does not exceed any of the FAP-associated levels.

\begin{figure*}
\centering
\includegraphics[width=0.9\textwidth]{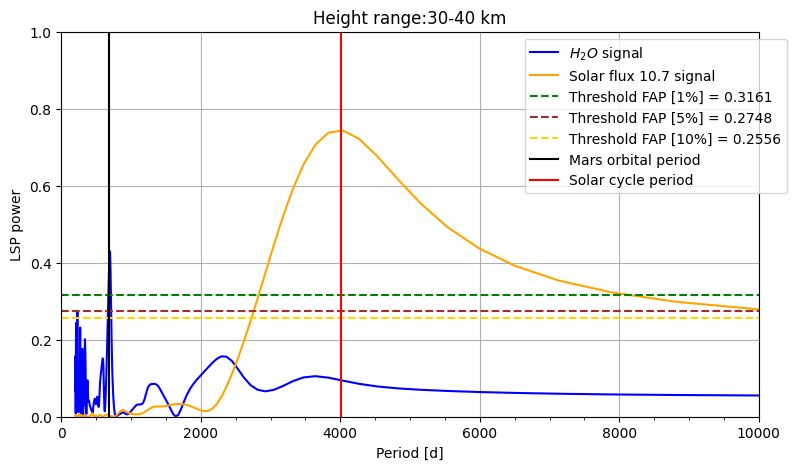}
\caption{Lomb-Scargle Periodogram for the altitude range 30-40 km.}
\label{fig:6:LSP_H2O_30-40}
\end{figure*}

\begin{figure*} 
\centering
\includegraphics[width=0.9\textwidth]{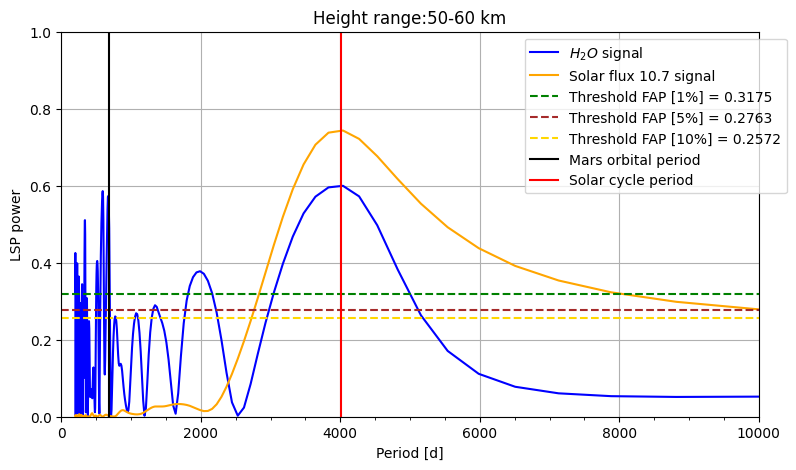}
\caption{Lomb-Scargle Periodogram for the altitude range 50-60 km.}
\label{fig:6:LSP_H2O_50-60}
\end{figure*}

Continuing with the analysis, in the altitude range of 50-60 km (see Figure \ref{fig:6:LSP_H2O_50-60}), it is noteworthy that one of the higher peaks (exceeding the FAP thresholds) appears to correspond to the solar activity period. This is evident both in the solar cycle time stamp and in the relationship with the solar activity frequency spectrum F 10.7 cm. For this same altitude range, the peak corresponding to the orbital period of Mars stands out above the noise threshold, along with a peak that seems to be centered at approximately 2000 days. We hypothesize that this peak may be linked to the same phenomenon responsible for generating peaks across all altitude ranges in the periodograms at this consistent periodicity.(approximately 2000 days).

\begin{figure*}
\centering
\includegraphics[width=0.9\textwidth]{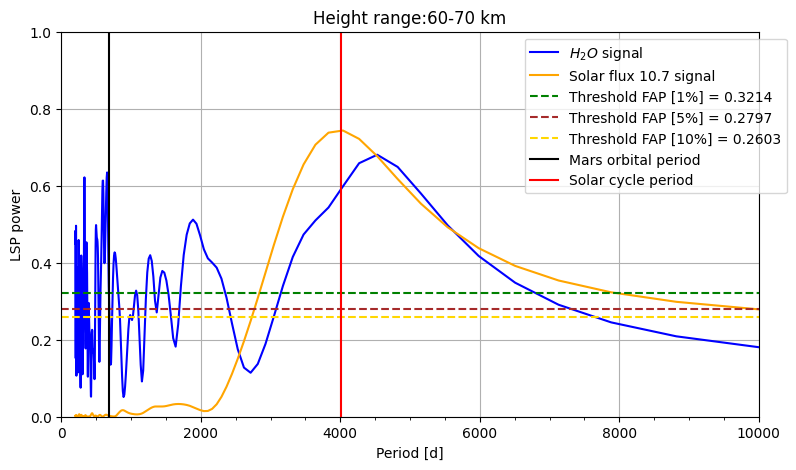}
\caption{Lomb-Scargle Periodogram for the altitude range 60-70 km.}
\label{fig:6:LSP_H2O_60-70}
\end{figure*}

In the periodogram corresponding to the altitude range 60-70 km (see Figure \ref{fig:6:LSP_H2O_60-70}), all the peaks of interest stand out: the peak associated with the orbital period on Mars, the pronounced peak at approximately 2000 days, and the peak at the solar cycle level. These three peaks are above the noise probability thresholds and stand out significantly in the frequency spectrum. A characteristic feature of the concentration of H$_2$O variation periodograms above 50 km in altitude is that the peaks associated with solar activity in these signals appear to be leading or lagging behind the solar activity period.\\

In light of the above and assuming the existence of a periodic link between the atmospheric H$_2$O concentration signal and the solar activity signal, a possible explanation is proposed: when studying the time period that encompasses a solar cycle, other periodic cycles are observed that occur at shorter time intervals, such as the orbital period of Mars. For these two cycles, there is a relationship between periods in days of 4015/687, which numerically results in a non-integer quantity. This physically implies that Mars' orbital phase does not coincide with the start or end of a solar cycle, and therefore, when the periodogram is calculated, this phase shift manifests as a slightly longer or slightly shorter period. However, there are moments when Mars receives solar maximum closest to its perihelion or aphelion. According to the ratio between periods, for an exact match between these two phases, Mars would need to complete more than 100 orbital periods to align with the same position where it coincided with a preceding solar maximum or minimum.

\begin{figure*}
\centering
\includegraphics[width=0.9\textwidth]{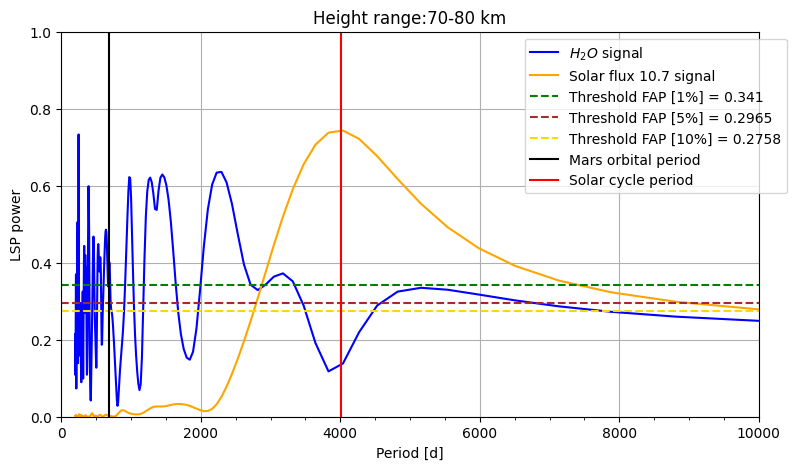}
\caption{Lomb-Scargle Periodogram for the altitude range 70-80 km.}
\label{fig:6:LSP_H2O_70-80}
\end{figure*}

\begin{figure*}
\centering
\includegraphics[width=0.9\textwidth]{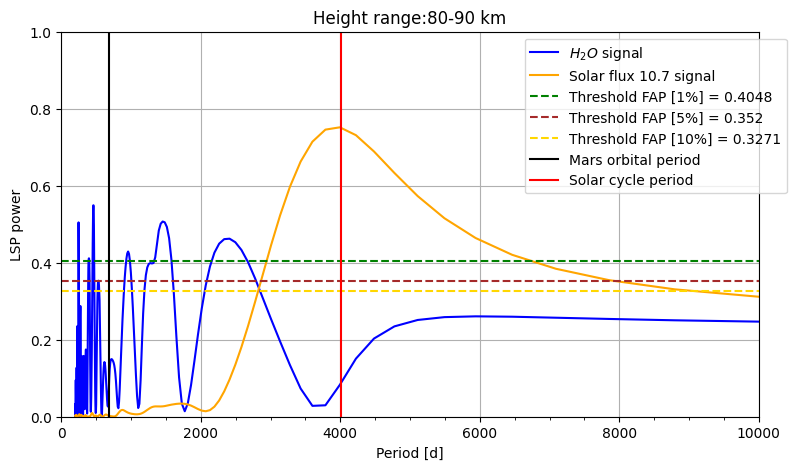}
\caption{Lomb-Scargle Periodogram for the altitude range 80-90 km.}
\label{fig:6:LSP_H2O_80-90}
\end{figure*}

Continuing with this analysis, the periodogram calculated for the altitude ranges 70-80 km (Figure \ref{fig:6:LSP_H2O_70-80}) and 80-90 km (Figure \ref{fig:6:LSP_H2O_80-90}) exhibit an anomalous behavior compared to the other periodograms inferred for the Martian mesosphere (between 50 and 100 km). The peak that could coincide with the solar cycle period in the 70-80 km range is a spurious peak (just exceeding the 5\% false alarm probability threshold) that appears to be shifted by 1000 days from the solar cycle time stamp. For the 80-90 km range, there is no peak associated with the solar activity cycle.

\begin{figure*}
\centering
\includegraphics[width=1\textwidth]{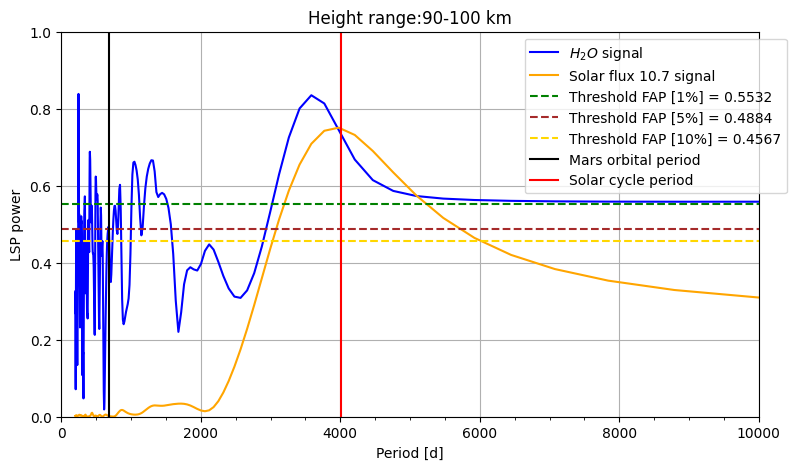}
\caption{Lomb-Scargle Periodogram for the altitude range 90-100 km.}
\label{fig:6:LSP_H2O_90-100}
\end{figure*}

Finally, for the 90-100 km range (see Figure \ref{fig:6:LSP_H2O_90-100}), the peak corresponding to the influence of solar activity is observed to be shifted by default. We assume that this shift is due to the reasons mentioned in previous paragraphs. In this particular range, the peak of the signal associated with the variability in H$_2$O concentration due to Mars' orbital period, while pronounced, does not exceed the second noise probability threshold (5\%). For this range, it is evident that the FAP levels exceed those of the previously sampled and analyzed inner layers. This may be due to the small amount of data retrieved from SPICAM on Mars Express for this altitude range, in addition to the high uncertainty associated with the derivations of these data.\\

In Figure \ref{fig:4:dispersion_all}, the distribution of observations for each altitude range is illustrated in detail. Here, the reader can analyze the abundance and dispersion of the data that affect the quality of the calculated periodograms. For example, in the case of the 90-100 km altitude range, there is a noticeable low volume of data compared to the other altitude ranges analyzed, and these data are associated with large uncertainties. These are the conditions that manifest in the high noise probability thresholds, which determine the limited statistical relevance of most of the signal obtained for this altitude range.\\

As mentioned earlier, there is a frequency peak in all the periodograms calculated for the different sampled altitude ranges, which falls between 2000 and 2250 days. For several altitude ranges (except for 20-30 km, 30-40 km, and 90-100 km), this peak has high statistical significance according to the FAP estimation method. This makes it plausible to raise new questions regarding the behavior of H$_2$O in the Martian atmosphere. Assuming that the periodograms provide reliable information about the periodic variability of atmospheric H$_2$O, what kind of phenomenon in the Martian atmosphere could be associated with a long period (between 2000 and 2250 days) that might have an impact on its periodic climatic variations? This is an open question that will be addressed later in the discussion prompted by the results obtained.\\

Although the most objective analysis to assess the reliability of peaks in a periodogram is the FAP, given the focus of this proposal, we decided to perform a standard deviation analysis for the characteristic peak of interest in the research, namely the peak associated with the periodic activity of the 11-year solar cycle. This analysis was conducted following the typical standard deviation method for signals with their corresponding SNR.\\

Firstly, the mean powers of the periodograms derived for the different altitude ranges and the mean powers of the errors of these signals were calculated. Using these values, the corresponding SNRs were calculated as the ratio between the average power of the two obtained signals. Then, Equation \ref{eq:5:sigma} \citep{LSP-Vardesplast} was implemented for all analyzed altitude intervals for the atmospheric chemical species in question (H$_2$O).

\begin{equation}
\centering
\label{eq:5:sigma}
\sigma_{\nu} = \nu_{\frac{1}{2}}\sqrt{\frac{2}{N\Sigma^2}}
\end{equation}

In this implementation process and its adaptation to the analysis of the calculated periodograms, it is important to clarify that the standard deviation over frequency ($\sigma_{\nu}$) has a standard deviation that must be related to periods through the inverse relationship between these two quantities. In this context, a method for error propagation over the period defined from the frequency must be implemented to obtain an estimate of the standard deviation that allows us to observe, in terms of the number of days, the confidence level of the peaks of interest in the periodograms. The classical model that relates period and frequency is $\nu_s = \frac{1}{T_s}$, and the propagated error over the period is:

\begin{equation}
\sigma T_s = T_s^{2}\sigma\nu_s    
\end{equation}

According to the model defined for the standard deviation ($\sigma_{\nu}$) in Equation \ref{eq:5:sigma}, the model implemented on the signals for the acquisition of standard deviations is:

\begin{equation}
\label{eq:6:error_prop_T}
\sigma T_s = \frac{T_s^{2}}{T}\sqrt{\frac{2}{N\Sigma^2}}    
\end{equation}

The synthesis of the standard deviation analysis can be seen in Table \ref{tab:6:errores_PLS_sigma}, where the SNR corresponding to each altitude range for the atmospheric chemical species H$_2$O in the region of interest are classified. It also expresses the calculated numerical period peak as well as the standard deviation of this value concerning the peak of interest in the research.

\begin{table}
\caption{Table: Standard Deviations and Errors in Days for the Peak of Interest}
\begin{equation}
\nonumber
\begin{array}{rrrr}
\text {Altitude Ranges [km]} & \text {SNR} & \Delta \mathrm{T} \text{ [days]} & \text{Error [\%]} \\
20-30 & 1664.03 & 0.15 & 0.004 \\
30-40 & 6358.03 & 0.04 & 0.001 \\
40-50 & 283.17 & 0.86 & 0.020 \\
50-60 & 55.61 & 4.39 & 0.103 \\
60-70 & 70.05 & 3.49 & 0.082 \\
70-80 & 1.29 & 189.82 & 4.473 \\
80-90 & 0.28 & 887.34 & 20.908 \\
90-100 & 1.85 & 133.45 & 3.144
\end{array}
\end{equation}
\label{tab:6:errores_PLS_sigma}
\end{table}

Note the small error levels for the altitude ranges 20-40 km, even when a peak exceeding the corresponding FAP levels is not pronounced. Therefore, it is important to consider that while standard deviation estimation is a common method for assessing the confidence of obtained results, it is not sufficient for LS-P analysis. A prior characterization of the statistical significance of a peak of interest through FAP threshold analysis is necessary. Then, if it surpasses the error thresholds according to the required level of precision, the deviation of the calculated peak should be studied. For our data, deviations ranging from 3.49 days for the 60-70 km altitude range concerning the maximum peak within the periodic limit of the solar cycle to 133.45 days for the 90-100 km altitude range are obtained.\\

\subsection{Comparing the atmospheres of Earth and Mars}

The purpose of the analysis presented below is to show how the coincidence of peaks in Solar flux variability occurs at higher altitudes for both Earth and Mars but not at lower altitudes. Since the data from Earth's atmosphere is from an empirical model, we can illustrate a juxtaposition between the expected outcome (Earth's case) and the obtained results (Mars's case).

For the data corresponding to the NRLMSISE-00 simulation at the low Earth ionosphere (105 km above the surface), concerning the chemical species that exhibited solar activity-related variability (N and O$_2$), periodic peaks in frequency spectra can be visualized in well-defined periodograms (with a higher curvature radius) compared to the frequency spectra peaks in the periodograms for the altitude ranges of 50-60 km, 60-70 km, and 90-100 km, which show variations in the concentration of H$_2$O in the Martian atmosphere. This might be due to the number of solar cycles encompassed in two different time windows, one for each data sample. For example, for the concentration distribution resulting from the NRLMSISE-00 simulation, there is a 60-year time window, covering approximately 5.46 solar cycles, in contrast to the time window for SPICAM detections from Mars Express, which corresponds to 14 years, encompassing only 1.28 solar cycle periods. This hypothesis is reinforced by observing the definition of frequency peaks in all calculated spectra for the periods corresponding to Earth's (NRLMSISE-00 data) and Mars's (SPICAM detections from MEx) orbital periods. In these two cases, 59 Earth orbital periods and 7.44 Martian orbital periods are covered.

\begin{figure*}
\centering
\includegraphics[width=1\textwidth]{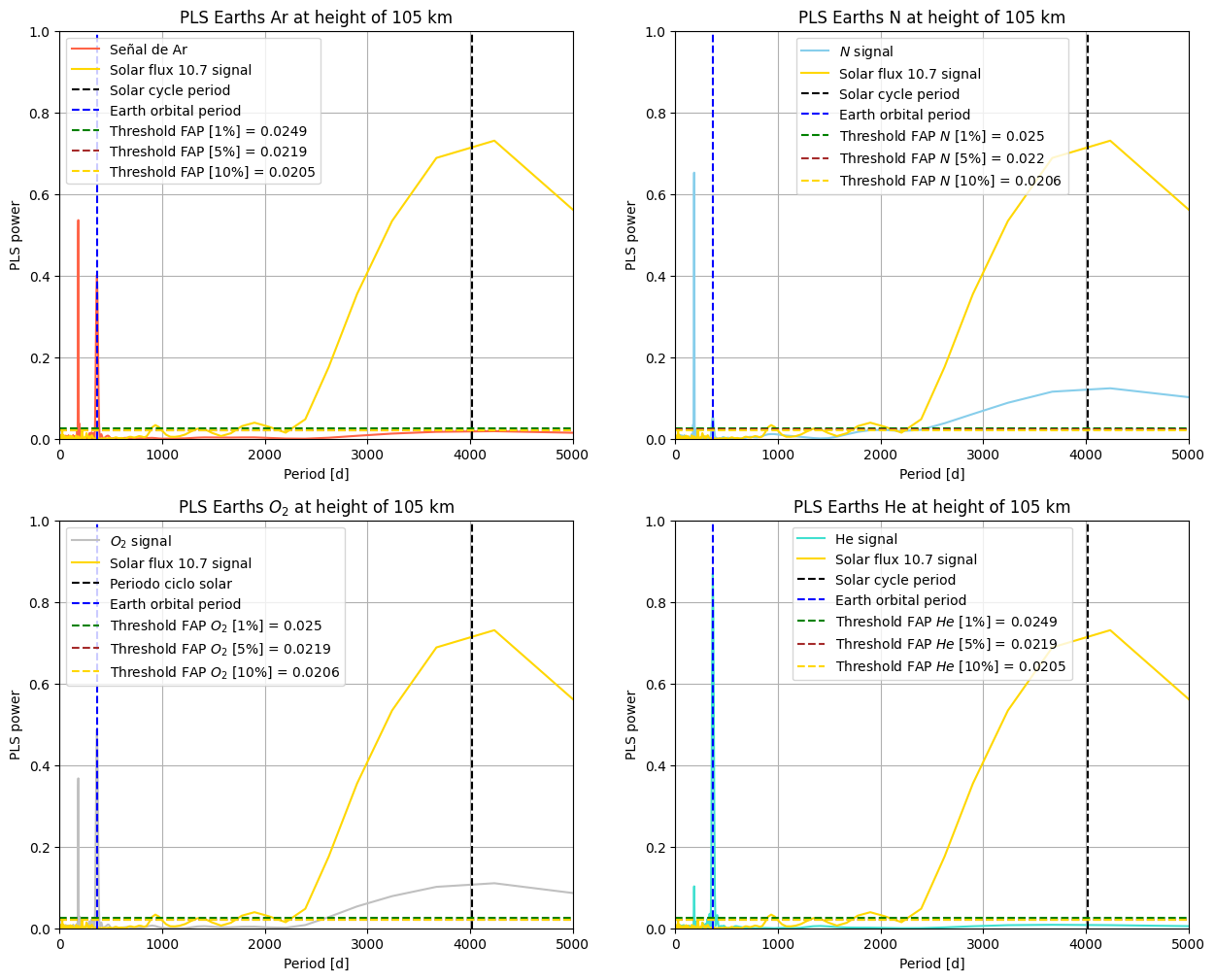}
\caption{Lomb-Scargle periodograms of different chemical species at 105 km altitude in Earth's atmosphere, in a 2004-2018 time window with a 20\% data volume reduction and uncertainties varying between 1\% and 10\% of the model estimates.}
\label{fig:6:LSP_earth_105_2004-2018}
\end{figure*}

The previous hypothesis is tested by calculating periodograms of data synthesized by NRLMSISE-00 for the time window corresponding to the time range of the Mars Express H$_2$O data (2004-2018). Additionally, the data set is randomly reduced to 20\%, creating a random systematic noise distribution varying between 1\% and 10\% regarding the model estimates. The resulting periodogram is shown in Figure \ref{fig:6:LSP_earth_105_2004-2018}.

One notable aspect of the periodograms calculated for the Earth case with a temporal resolution corresponding to the 2004-2018 time window is that they reach very low power levels compared to the Martian signal peaks. This may be because, although there is an influence of solar activity on the concentrations of the sampled chemical species (as seen in the periodograms for a 60-year time grid from 1961 to 2021 in Figure \ref{fig:5:Earth_105_LSP}), this influence is less pronounced for Earth due to the significant difference in densities between Earth's and Mars's atmospheres \footnote{The average density of Mars' atmosphere is 1\% of Earth's atmosphere density.}. The heights of the peaks in the Martian atmosphere layers are higher than those corresponding to the peaks in Earth's 105 km altitude, even though the time window for Earth's data is over 4 times longer. In summary, Mars' lower atmosphere is more strongly influenced by solar activity than Earth's lower atmosphere.

The reader may notice that the signals in the new calculated periodograms show a greater number of very faint peaks. This is because the implemented algorithm of the classical LS-P model identifies non-uniform gaps generated by randomly reducing the data as points through which harmonic functions with random frequencies could pass, reflecting in the corresponding frequency spectrum.

Finally, while our method illustrates the influence of solar activity on planetary atmospheres quite well, it does not allow us to quantify it in a way that establishes a numerical equivalence between the strength of the effect triggered in one or the other studied atmosphere.

\section{Discussion and conclusions}

The purpose of the following analysis is to investigate a possible relationship between the periodic behavior of solar activity using the characteristic 10.7 cm radio flux proxy and variations in the concentration of H$_2$O in different layers of the Martian atmosphere. The results obtained indicate that for certain layers of the high mesosphere of Mars, there is a connection between solar activity and variability in atmospheric H$_2$O concentration. This relationship is revealed in the periodograms of both signals, showing a statistically significant power peak around the 11-year solar cycle timestamp according to the LS$-$P method.

While the variability on Mars may not be directly driven by this radio flux indicator, strong findings lead to the conclusion of such a relationship. The initial implementation of the LS$-$P method passed the effectiveness test by generating periodograms for a series of concentration signals in Earth's lower ionosphere at an altitude of 105 km, revealing peaks at the characteristic solar cycle frequency of $\frac{1}{4015}\approx 2.49\times10^{-4}$ d$^{-1}$ that were not present at an altitude of 55 km above the Earth's surface. In this case, the data was generated by the empirical simulator NRLMSISE$-$00 and, therefore, had no associated direct error. Information was abstracted from the simulation, which, while modeled under mathematical parameters and adjustments according to known atmospheric physics \citep{hedin1991extension_NRLMSISE,walker1965analytic}, had not been represented in this way before.

The successful representation of variability signals in the concentration of various Earthly chemical species allowed the analysis of frequency spectrum synchrony with expected periodic activities such as variability associated with Earth's seasonal cycles, for both altitudes simulated (55 and 105 km). The alignment of frequency peaks with the temporal marks of these variable activities enabled us to refer to the method's utility for analyzing the signal of interest in the research: H$_2$O concentration in the Martian atmosphere.

Our findings reveal a relationship between periods of H$_2$O concentration variability in the Martian mesosphere (50-100 km) and solar activity, consistent with previous studies such as \cite{fedorova2018water,fedorova2009solaroccult}. These studies reveal concentration changes due to different seasonal stages on Mars, which are summarized in our analysis in the significant power peak in almost all calculated periodograms (altitude range 20-90 km) around the Martian orbital period.\\

The interpretation of our results, summarized in the periodograms calculated for different altitude ranges in the confined region of Mars: between 60$^\circ$-80$^\circ$ N latitude and 0$^\circ$-80$^\circ$ longitude (see Figure \ref{fig:4:Marte_distrib}), suggests that the incorporated method serves as an effective technique for studying planetary atmosphere data. The Lomb-Scargle periodogram appears, in light of the evidence presented in this research, as a powerful algorithm for extracting information associated with the periodic behavior of various phenomena in planetary atmospheres. The main aspect that makes this method so important for such studies is that it does not require a sample of data evenly distributed in a time grid. This is a common feature of data obtained from planetary orbiters, which collect spectral samples from different regions as they orbit, with their orbits not homogeneously synchronized with planetary regions unless their orbits remain stationary with respect to the planet's rotation period.

Although direct causation associated with the relationship between periods of activity in the two studied signals cannot be conclusively established, some hypotheses can be postulated. The radio flux indicator (solar activity signal) is correlated with EUV and UV fluxes, as seen in the work of \cite{bruevich2011solar_corr_10.7_wolf}, as Figure \ref{fig:1:pectinton_cosmic} illustrates. These indicators, in turn, are correlated with solar wind intensity and SEP (Solar Energetic Particles) abundance \citep{malandraki2018_SEP}, which are energetic particles that carry and deposit energy in the form of heat in the high ionosphere of planetary atmospheres. Energy transport phenomena, along with the internal atmospheric dynamics of the planet, could lead to the warming of internal regions in correlation with the intensity and quantity of traveling SEP, modifying the H$_2$O concentration at different altitudes in the Martian atmosphere, including internal regions as shown in the presented research.

The appearance of a remarkable power peak, completely unexpected, at a period of approximately 2250 days, in various signals of variability of the chemical species of interest on Mars, demonstrates the ability of the implemented method to study new signals of variability over different time ranges. Due to research interests, the upper-frequency limit was set to correspond to a 20-day period. However, this does not imply that there are no signals that can be detected beyond this range, possibly related to phenomena characterized by the variable activity of the Martian atmosphere. For example, it is known from models that GWs are below 2.9 MHz \citep{astafyeva2019ionosphericWG}, and therefore, they could stimulate the appearance of oscillatory phenomena encompassing frequencies below this threshold or reinforce them through constructive interference with other oscillatory phenomena of similar frequencies. An example of a long period to note in this regard is the Quasi-Biennial Oscillation (QBO) which is driven by the periodic interaction of eastward and westward winds in the stratosphere in the Earth's tropical region, with an average period of 28 to 29 months. According to recent studies, QBO could be driven by atmospheric GW
\citep{takahashi1996simulation_QBO,giorgetta2002forcing_QBO,scaife2000realistic_QBO}. These periodic variations were not observed in our periodograms obtained for Earth's atmospheric data in any of the sampled observables. This may be due to the fact that such long-period oscillations do not stimulate variable activity in the concentration of these chemical species (Ar, He, N, O$_2$), or, more likely, the NRLMSISE-00 model does not take into account these oscillatory phenomena, which are still a subject of strong debate in the scientific community. Another essential aspect in this regard is that our variability data was simulated for 2 points in the three-dimensional Earth's atmosphere with latitude, longitude, and altitude coordinates (55, 45, 55) and (55, 45, 105), not layers or spatial grids. Perhaps for some points in the Earth's atmosphere (near the equator for QBO), these changes could manifest themselves, making this another possible research path from simulations in the NRLMSISE$-$00 model: the connection between variability at different points in the Earth's atmosphere, stimulated by the variable activity of points in other regions. The QBO may be a suitable starting point for this; there could be a manifestation in Martian climate similar to oscillatory phenomena like Earth's QBO and could be explored by working with well-known Martian climate simulations such as Martian Atmospheric Waves Perturbation Datasets \citep[MAWPD,][]{zhang2023observation_WG_dataset}. Although research in these fields is still in its early stages, works like ours could open the doors to new questions that, if addressed, would lead to expanding the current knowledge path of atmospheric physics.\\

Finally, it is worth noting that this study of variability was conducted on one observable (H$_2$O concentration in the Martian atmosphere) associated with variable activity of the solar cycle. This study was part of a set of multiple known observables, including concentrations of various neutral chemical species, concentrations of different ionized species, magnetic fields, temperatures, wind speeds, and more. It was also conducted over a time grid comparable to the 11-year solar cycle period of interest. The results obtained were favorable and pave the way for adapting this method to study various oscillatory observables associated with known periodic phenomena, such as seasonal variation, QBO, solar cycle, Earth's oceanic oscillations (El Niño and La Niña), and others, for different planetary environments.

\section*{Acknowledgements}

JNMC would like to extend special thanks to the Group of Solar Astrophysics (GoSA) and the Observatorio Astronómico Nacional de Colombia for providing the space where the various feedback received greatly guided this research. We also extend our gratitude to James R. Murphy and Ana Fedorova for their advice on processing data from older Mars orbiters such as Mars Express and Mars Reconnaissance Orbiter (MRO), and their work that has served as the foundation for the development of this research. Finally, a heartfelt thanks to the Orbitamautas Research Group for providing an informal space where the communicative capacity of this work could be put to the test. \\

\section*{Data Availability}

The entire implementation, adaptation, and data analysis procedure is developed in the Python programming language, \url{https://www.python.org}. The method is coded for each database in a Jupyter notebook in the GitHub: repository:: \url{https://github.com/NicolasLebrum/Planetary-Atmospheres.git}.



\bibliographystyle{mnras}
\bibliography{references} 




\bsp	
\label{lastpage}
\end{document}